\documentclass[aps,prd,reprint,nofootinbib,superscriptaddress,floatfix]{revtex4-2}

\usepackage{amsmath,amssymb,amsfonts}
\usepackage{empheq}
\usepackage{graphicx}
\usepackage{hyperref}
\usepackage{mathtools}
\usepackage{xcolor}
 
\def\Tr{\mathop{\mathrm{Tr}}}
\newcommand{\dd}{\mathrm{d}}

\begin{document}

\title{Accurate pulsar timing array likelihoods: better low-rank covariance for gravitational wave background}

\author{Heling Deng}
\affiliation{Columbia Astrophysics Laboratory, Columbia University, New York, NY 10027, USA}

\author{Ken D. Olum}
\affiliation{Institute of Cosmology, Department of Physics and Astronomy, Tufts University, Medford, MA 02155, USA}

\author{Rutger van Haasteren}
\affiliation{Max Planck Institute for Gravitational Physics (Albert Einstein Institute), D-30167, Hannover, Germany}
\affiliation{Leibniz Universität Hannover, D-30167, Hannover, Germany}

\begin{abstract}

In pulsar timing array analyses, the stochastic process describing the gravitational wave background is typically modeled using a Fourier basis at discrete frequencies with an interval $\sim 0.1~\mathrm{yr}^{-1}$, with a diagonal Fourier-domain covariance determined by the power spectral density at the corresponding frequencies. Although this rank-reduced modeling significantly accelerates likelihood evaluation and provides a good approximation in many scenarios, it can lead to biased parameter inference because it is insensitive to spectral features on scales smaller than $\sim 0.1~\mathrm{yr}^{-1}$. We introduce a better construction of the low-rank covariance for low-frequency processes by maximizing the log-likelihood averaged over an ensemble of realizations drawn from the true model likelihood. In this approach, the basis-domain covariance matrix is sensitive to the entire model spectrum and is generally non-diagonal. Using two mock datasets, we demonstrate improved likelihood accuracy and more faithful recovery of spectral features compared to the conventional Fourier-diagonal model. Benchmark tests show that this new model requires a computational cost comparable to that of the conventional approach.
\end{abstract}
\maketitle

\section{Introduction}

Pulsar timing arrays (PTAs) have reported strong evidence for a stochastic gravitational wave background (GWB) in the nanohertz frequency band \cite{NANOGrav:2023gor,Reardon:2023gzh,Xu:2023wog,EPTA:2023fyk,miles2025meerkat}. As PTA observations grow in duration and incorporate an increasing number of precisely timed millisecond pulsars, the statistical modeling of low-frequency processes, including the GWB, intrinsic achromatic pulsar noise, and chromatic noise induced by the environment, has become central to data analysis. Accurate likelihood evaluation is essential not only for robust characterization of the GWB spectrum, but also for hypothesis testing, model comparison, and searches for potential individual signals, such as continuous waves from supermassive black hole binaries.

Typically, PTA data are modeled with a multivariate normal likelihood whose covariance matrix encodes contributions from both white noise and low-frequency processes. Assuming that the low-frequency processes are stationary Gaussian, the covariance $C_{\rm gp}$ is determined via the Wiener-Khinchin theorem. After marginalizing over timing model parameters, the resulting marginalized likelihood involves a projected covariance matrix \cite{vanHaasteren:2012hj}. While this formulation is exact, its direct evaluation is computationally prohibitive: the dimension of $C_{\rm gp}$ is set by the total number of times of arrival (TOAs) of the pulses from pulsars, typically $\sim 10^{5}\text{--}10^{6}$ in modern PTA datasets. Constructing, inverting, and computing the determinants of such a large matrix is infeasible for practical Bayesian inference.

To address this challenge, almost all contemporary PTA analyses adopt a low-rank approximation to $C_{\rm gp}$. The most widely used approach represents the processes as a truncated Fourier expansion with a finite number of low-frequency modes, whose coefficients are modeled as independent Gaussian variables \cite{Lentati:2012xb}. This results in a covariance of the form $C_{\rm gp} \approx F \Phi F^\top$, where $F$ is a Fourier design matrix, and the matrix $\Phi$ is diagonal, encoding the power spectral density (PSD) at corresponding discrete frequencies. Combined with the Woodbury identity, this approximation leads to a PTA likelihood that involves matrices whose dimension is the number of Fourier basis functions, typically $\sim 10$--$100$. This framework is highly effective in the analyses of real PTA datasets.

Despite its computational efficiency, this Fourier-diagonal approximation has intrinsic limitations. Because the PSD is evaluated only at discrete frequencies separated by $1/T$, where $T$ is usually the total observation span, the model is insensitive to spectral features on finer scales, including those at frequencies below $\sim 1/T$, which may still leave an imprint in the timing residuals after the timing model marginalization. More generally, the Fourier-diagonal approach cannot see any spectral structures supported only within an interval of width $1/T$ between frequencies.  If the correct spectrum has such a feature, this approach rejects it in favor of a broader feature that overlaps the frequencies that it uses. These effects can lead to inaccurate likelihood evaluations and, consequently, biased parameter inference.

In this work, we address this issue by revisiting the construction of the low-rank covariance for the low-frequency processes from a statistical perspective. We emphasize that the covariance matrix enters PTA analyses only through the likelihood, which defines a probability distribution over the timing residuals. We therefore seek the low-rank model that optimally reproduces the ensemble of residuals drawn from the true model, rather than one that merely matches the covariances. Concretely, we derive $\Phi$ by maximizing the ensemble average log-likelihood. This procedure gives an optimal solution for $\Phi$, which we refer to as the MEAL (\emph{maximizing ensemble average (log-)likelihood}) covariance.

The MEAL covariance includes off-diagonal correlations in the basis domain and accounts for both timing model projection and white noise weighting. Importantly, this new framework is sensitive to spectral features below $1/T$ and within narrow frequency bands without increasing the number of basis functions.

In Sec.~\ref{SecII}, we review the likelihood of timing residuals and describe the low-rank approximation widely adopted in current PTA analyses, along with its limitations. In Sec.~\ref{SecIII}, we present the construction of the MEAL model in detail, from both theoretical and numerical perspectives. In Sec.~\ref{SecIV}, we demonstrate the performance of the method using two mock datasets. Finally, we summarize and discuss our conclusions in Sec.~\ref{SecV}. We discuss several technical issues and
alternative formulations in Appendices \ref{appendA}--\ref{appendD}.

\section{Ideal PTA likelihood and conventional approximation \label{SecII}}
In this section, we briefly review the PTA likelihood and explain how the low-rank approximation for the covariance of low-frequency noise processes significantly accelerates the likelihood evaluation. We also describe the conventional Fourier-based model widely adopted in PTA analyses, discuss its limitations, and summarize some attempts to improve it.

\subsection{Ideal model}
Pulsars are rapidly spinning, highly magnetized neutron stars that emit beams of radio waves sweeping across Earth. Millisecond pulsars are known to have remarkably stable rotations, allowing the TOAs to be measured with extraordinary precision. The observed TOAs are well described by a deterministic timing model that accounts for the pulsar’s spin-down, orbital motion, the propagation of radio pulses through the interstellar medium, etc.  After subtracting the best-fit timing model, the resulting timing residuals typically consist of white noise from measurement uncertainties, small perturbations to the timing model, and low-frequency processes (with the timing model components removed). For simplicity, we assume that the low-frequency processes are achromatic (independent of the observation radio frequency), including intrinsic pulsar noise and the GWB. Extending the analysis to chromatic noise, such as dispersion measure variations, is straightforward and is discussed in Appendix~\ref{appendD}.

Assuming that the low-frequency processes are stationary Gaussian, the log-likelihood of the residuals is given by \cite{vanHaasteren:2008yh}
\begin{equation}
\log\mathfrak{L}=-\frac{1}{2}\left[\left(r-M\epsilon\right)^{\top}C^{-1}\left(r-M\epsilon\right)+\log\det \left(2\pi C\right)\right], \label{vanHaasterenLevin}
\end{equation}
where $r$ is the vector of residuals for all pulsars in the array,
$C$ is the total covariance matrix for both white noise and low-frequency Gaussian processes, $M$ is the timing model design matrix, and $\epsilon$ denotes small deviations from the best-fit timing model parameters. The covariance matrix can be decomposed as
\begin{equation}
C=N+C_{\rm gp},
\end{equation}
where $N$ is the white noise covariance, whose parameters are usually fixed from individual pulsar analyses, and $C_{\rm gp}$ is computed via the Wiener-Khinchin theorem:
\begin{equation}
\left(C_{\rm gp}\right)_{ij}=\int_{0}^{\infty}\!S(f)\cos\!\left[2\pi f(t_{i}-t_{j})\right]\dd f,\label{eq:WK}
\end{equation}
where $t_{i}$ is the $i$-th TOA, and $S(f)$ is the PSD of the low-frequency processes. In practice, $S(f)$ is often parameterized either by a simple functional form, such as a power law, or by an expression determined by underlying physical parameters. Therefore, the PTA likelihood is generally a function of the PSD parameters, including those that characterize the GWB.

Since the TOAs are dominated by the timing models, in Bayesian inference we can assign uninformative priors to the timing model parameters, and thus to $\epsilon$. This allows us to integrate out $\epsilon$ analytically. Up to a constant prefactor, the resulting marginalized PTA likelihood $\mathcal{L}=\int\mathfrak{L}\dd\epsilon$ takes the form of a multivariate normal distribution \cite{vanHaasteren:2012hj}:
\begin{equation}
\log\mathcal{L}=-\frac{1}{2}\left[r^{\top}GC_G^{-1}G^{\top}r+\log\det\left(2\pi C_G\right)\right],\label{eq:trueL}
\end{equation}
where the columns of $G$ are orthonormal and span the subspace orthogonal to the column space of $M$ (so $G$ plays the role of projecting out the timing model), and $C_G \equiv G^\top C G$ can be interpreted as the projected covariance.

In principle, Bayesian inference can be performed using this exact likelihood, sampling over the PSD parameters. In practice, however, evaluating Eq.~(\ref{eq:trueL}) is computationally prohibitive: the dimension of $C$ is the total number of TOAs, $N_{\rm toa} \sim 10^{5}\text{--}10^6$. The projected covariance $C_G$ has a smaller dimension, but only by a number comparable to the number of timing model parameters, typically $\sim 10\text{--}100$. Constructing, inverting, and taking the determinant of such a large matrix is extremely costly.

It is useful to obtain the marginalized likelihood $\mathcal{L}$ in an equivalent form by imposing an improper Gaussian prior on $\epsilon$ \cite{NANOGrav:2023icp}:
\begin{equation}
    \mathcal{L}=\int \mathfrak{L} \frac{\exp\left(-\frac{1}{2}\epsilon^\top E^{-1}\epsilon\right)}{\sqrt{\det\left(2\pi E\right) }}\dd \epsilon, 
\end{equation}
where $E$ is a diagonal covariance matrix with infinite variances. The result is \cite{vanHaasteren:2008yh,vanHaasteren:2014qva}
\begin{align}
    \log \mathcal{L} =& -\frac{1}{2} r^{\top}\left(C+MEM^\top\right)^{-1}r\nonumber\\
    &-\frac{1}{2}\log\det\left[2\pi \left(C+MEM^\top\right)\right]. \label{eq:trueL2}
\end{align}
We next review the standard low-rank approximation that makes this likelihood fast to evaluate.

\subsection{Low-rank approximation}

A common strategy to handle the formidable $C_{\rm gp}$ is to approximate it with a low-rank representation. For clarity, we first consider a single pulsar and its intrinsic noise; extending to a common process across all pulsars with pulsar-pair correlations is straightforward and will be discussed in Subsection~\ref{HD}. We approximate $C_{\rm gp}$ by (in what follows, tildes denote approximate quantities):
\begin{equation}
C_{\rm gp}\approx\tilde{C}_{\rm gp}\equiv F\Phi F^{\top},
\end{equation}
where $F$ is an $N_{\rm toa} \times N_F$ matrix whose columns form a set of basis functions, and $\Phi$ is a symmetric $N_F \times N_F$ matrix describing $\tilde{C}_{\rm gp}$ in the basis domain. 

By Eq. \eqref{eq:trueL2}, the approximate likelihood becomes \cite{NANOGrav:2023icp}
\begin{equation}
\log\tilde{\mathcal{L}}  =-\frac{1}{2}\left[r^{\top}K^{-1}r+\log\det \left(2\pi K\right)\right], \label{eq:Lapprox}
\end{equation}
where
\begin{equation}\label{eq:Kwhere}
K=D+F\Phi F^\top
\end{equation}
with
\begin{equation}\label{eq:Ddef}
    D=N+MEM^\top.
\end{equation}
The Woodbury matrix identity gives
\begin{equation}
K^{-1}=D^{-1}-D^{-1}F\left(I+ \Phi F^{\top}D^{-1}F\right)^{-1}\Phi F^{\top}D^{-1}, \label{eq:L_diagphi}
\end{equation}
with
\begin{align}
D^{-1}&=N^{-1}-N^{-1}M\left(M^{\top}N^{-1}M\right)^{-1}M^{\top}N^{-1}\nonumber\\
&= G\left(G^\top N G\right)^{-1}G^\top,\label{eq:D}
\end{align}
where we have used $E^{-1}=0$. Here, $D^{-1}$ can be interpreted as the white noise precision matrix projected onto the subspace orthogonal to the timing model. There are several advantages to introducing $\Phi$ and writing the likelihood in the form of Eq.~\eqref{eq:Lapprox}. Since $N$ and $M$ are fixed before the inference, $D^{-1}$, and thus $r^\top D^{-1} r$, $F^\top D^{-1} r$, and $F^\top D^{-1} F$ can be precomputed. The matrix inverse that cannot be precomputed, $\left(I+ \Phi F^{\top}D^{-1}F\right)^{-1}$, has dimension $N_F$, and is computationally much cheaper than $C^{-1}_G$. Moreover, the matrix determinant lemma gives
\begin{equation}\label{eq:detK}
\det K = \det \left(I+ \Phi F^{\top}D^{-1}F\right)\det D
\end{equation}
and we can precompute
\begin{equation}\label{eq:detD}
\det D = \det \left(M^\top N^{-1}M\right)\det N\det E.
\end{equation}
Here $\det E$ is formally infinite, but we can just leave it out since $E$ is a constant.

A variety of choices are available for the basis design matrix $F$. Global bases, such as Fourier modes and Legendre polynomials, can efficiently capture low-frequency signals, but they couple all data points together and may be less effective for irregular cadences or gapped data. Furthermore, truncating a global basis can cause oscillatory behavior known as the Gibbs phenomenon \cite{Wilbraham1848,Gibbs1899}, analogous to the Runge phenomenon in polynomial approximation \cite{Runge1901}. This effect was recently discussed in the PTA context in Ref. \cite{Crisostomi:2025vue}. Local bases, such as B-splines and Gaussian radial basis functions, are more flexible because their support can be adapted to irregular cadences.

Once the basis $F$ is fixed, what should we choose for $\Phi$? Different choices of $\Phi$ can give different approximations to $C_{\rm gp}$ even for the same basis. We first review the conventional choice in the next subsection.

\subsection{Current standard: Fourier basis \texorpdfstring{$F$}{F} and diagonal~\texorpdfstring{$\Phi$}{Phi}}

Beginning with Ref. \cite{Lentati:2012xb}, the standard practice in PTA analysis has been to take the basis $F$ as a Fourier design matrix with a high-frequency cutoff at the $n_{F}$th Fourier mode:
\begin{widetext}
\begin{equation}
F_{\rm Fourier}=\begin{pmatrix}\sin\!\left(2\pi f_{1}t_{1}\right) & \cos\!\left(2\pi f_{1}t_{1}\right) & \cdots & \sin\!\left(2\pi f_{n_{F}}t_{1}\right) & \cos\!\left(2\pi f_{n_{F}}t_{1}\right)\\
\sin\!\left(2\pi f_{1}t_{2}\right) & \cos\!\left(2\pi f_{1}t_{2}\right) & \cdots & \sin\!\left(2\pi f_{n_{F}}t_{2}\right) & \cos\!\left(2\pi f_{n_{F}}t_{2}\right)\\
\vdots & \vdots & \vdots & \vdots & \vdots\\
\sin\!\left(2\pi f_{1}t_{N_{\rm toa}}\right) & \cos\!\left(2\pi f_{1}t_{N_{\rm toa}}\right) & \cdots & \sin\!\left(2\pi f_{n_{F}}t_{N_{\rm toa}}\right) & \cos\!\left(2\pi f_{n_{F}}t_{N_{\rm toa}}\right)
\end{pmatrix},\label{Fourier_basis}
\end{equation}
\end{widetext}
where $f_{k}=k/T$, with $k=1,2,\ldots,n_{F}$ and $T$ usually being either the pulsar's or the whole PTA's observation span,\footnote{For example, in the fiducial GWB model of NANOGrav's 15-year data analysis \cite{NANOGrav:2023gor}, $n_F=30$ for intrinsic pulsar noise, $n_F=14$ for GWB, and $T=16.03$ yr is the time between the first and last observations in the whole dataset. For individual pulsar timing \cite{NANOGrav:2023hde} and noise analyses \cite{NANOGrav:2023ctt}, however, $T$ was chosen to be each pulsar's observation span.} Each Fourier mode has a sine component and a cosine component, so $n_F=N_F/2$. The matrix $\Phi$ is taken to be diagonal, encoding the PSD:
\begin{equation}
\Phi_{\rm diag}=\text{diag}\!\begin{pmatrix}\rho_{1} & \rho_{1} & \cdots & \rho_{n_{F}} & \rho_{n_{F}}\end{pmatrix},\quad \rho_{k}=S(f_{k})\Delta f,\label{eq:DiagPhi}
\end{equation}
where $\Delta f=1/T$ is the difference between adjacent frequencies. The likelihood in Eq. \eqref{eq:Lapprox}, with $F=F_{\rm Fourier}$ and $\Phi=\Phi_{\rm diag}$, which is similar in spirit to the (truncated) Whittle likelihood \cite{whittle1951hypothesis}, is implemented in production-level GW search pipelines, such as \texttt{Enterprise} \cite{enterprise} and \texttt{Discovery} \cite{Vallisneri_nanograv_discovery_2025}. 

Writing $\tilde{C}_{\rm gp}=F\Phi_{\rm diag} F^{\top}$ explicitly gives
\begin{equation}
\left(\tilde{C}_{\rm gp}\right)_{ij}= \Delta f\sum_{k=1}^{n_{F}}S(f_{k})\cos\!\left[2\pi f_{k}(t_{i}-t_{j})\right],\label{eq:discreteWK}
\end{equation}
which is immediately recognizable as a discrete approximation to the Wiener-Khinchin integral \eqref{eq:WK}, with $n_F \Delta f$ being the upper bound. Alternatively, this method can be understood as modeling a process that emits only at the discrete frequencies $f_k$, with $\Phi_{\rm diag}$ as the covariance of the Fourier coefficients at those frequencies. Thus, $\Phi_{\rm diag}$ being diagonal means the coefficients are assumed to be uncorrelated.

The accuracy of this modeling depends on how much the PSD varies within the frequency interval $\Delta f$. For smooth spectra such as a power law ($S(f)\propto f^{-\gamma}$), Eq.~\eqref{eq:discreteWK} may be an acceptable approximation. However, because the PSD is evaluated only at discrete frequencies $f_{k}=k\Delta f$, spectral features on scales finer than $\Delta f$ are ignored, including those at $f<1/T$. Although most of the power at such low frequencies is absorbed by the timing model fit since a sinusoid with $f<1/T$ is highly covariant with the pulsar's quadratic spin-down, some information may still remain.

If we decrease $\Delta f$ and increase $n_F$ to cover the same frequency range, the discrete sum in Eq.~(\ref{eq:discreteWK}) converges to the Wiener-Khinchin integral. For example, we can get a better approximation $\tilde{C}_{\rm gp}$ with $\Delta f = 1/(10T)$ and $n_F\to10n_F$ \cite{Zhang:2026qqr}. An alternative strategy is to evaluate the sum using logarithmically spaced frequency components rather than uniform spacing to better capture the low-frequency behavior of the PSD \cite{vanHaasteren:2014faa}. However, these approaches come at the cost of increasing the dimensionality of $\Phi_{\rm diag}$.

\subsection{Attempts at improvement: matching $C_{\rm gp}$ and $\tilde{C}_{\rm gp}$ \label{matchingCandC}}

Rather than refining the frequency grid with spacing $\Delta f$ for the diagonal of $\Phi$, can we obtain a better approximation by exploiting the off-diagonal entries of $\Phi$, without increasing its dimension? After all, a major computational bottleneck in likelihood evaluation is the matrix inverse $\left(I+ \Phi F^{\top}D^{-1}F\right)^{-1}$, whose dimension is $N_F$.

Recently, Ref. \cite{Allen:2024uqs} (using a Fourier basis) and Ref. \cite{Allen:2025waa} (using a Legendre polynomial basis) derived expressions for $\Phi$ that lead to the exact match $C_{\rm gp}=F\Phi F^\top$.\footnote{An advantage of the Legendre polynomial basis considered in Ref. \cite{Allen:2025waa} is that the first three components can be removed a priori, as they do not contribute to the post-fit residuals after marginalizing over the timing model perturbations.} Both derivations assume continuous observing time over a finite observation span. The exact equality requires a complete and infinite set of basis functions, and hence a $\Phi$ with infinite dimensions. One may then construct an approximate covariance $\tilde{C}_{\rm gp}$ by truncating $F$ and $\Phi$. 

From a more practical perspective, Ref. \cite{Crisostomi:2025vue} proposed a $\Phi$ that is a coarse version of the true $C_{\rm gp}$ computed at a set of ``nodes'' distributed over the observation span. This $\Phi$ can be obtained efficiently using the Wiener-Khinchin theorem via a fast Fourier transform (FFT). To expand this coarse covariance into $\tilde{C}_{\rm gp}=F \Phi F^\top$, one may choose $F$ to be the linear interpolation design matrix that connects the nodes. This model is referred to as ``FFT-Int''.

Another way to match $C_{\rm gp}$ and $\tilde{C}_{\rm gp}$ is to consider the ``distance'' between the two matrices. A standard measure is the Frobenius norm $\left\Vert C_{\rm gp}-\tilde{C}_{\rm gp}\right\Vert $. Minimizing the norm with respect to $\Phi$ gives
\begin{equation}
\Phi=\left(F^{\top}F\right)^{-1}\!\left(F^{\top}C_{\rm gp}F\right)\!\left(F^{\top}F\right)^{-1}.\label{eq:matchC}
\end{equation}
If $F=F_{\rm Fourier}$, this $\Phi$, by design, results in a better approximation $\tilde{C}_{\rm gp}$ than $\Phi_{\rm diag}$ does. If $F$ is chosen to be the linear interpolation design matrix, this $\Phi$ should give a better $\tilde{C}_{\rm gp}$ than that obtained by the FFT-Int method described above.

Can we do even better? Note that reproducing $C_{\rm gp}$ is not the ultimate goal. The observed data, or more precisely, the timing residuals, represent a single realization drawn from the true likelihood $\mathcal{L}$. What we really need is an approximate distribution $\tilde{\mathcal{L}}$ that best generates the ensemble of such realizations. What is the $\Phi$ that gives this optimal $\tilde{\mathcal{L}}$?

\section{Projected basis and likelihood-optimized covariance \label{SecIII}}

In this section, we show that a better choice of $\Phi$ can be found by working directly with the likelihood. We will also provide an efficient way to obtain it numerically. 

\subsection{Maximizing ensemble average log-likelihood}

The likelihood $\mathcal{L}$ describes a normal distribution from which the ``projected residuals'' $G^\top r$ are drawn. Consider a large ensemble of realizations generated from the true model and evaluate their average log-likelihood under the approximate model:
\begin{equation}
\left\langle \log\tilde{\mathcal{L}}\right\rangle  
  =-\frac{1}{2}\left[\left\langle r^{\top}G\tilde{C}_{G}^{-1}G^{\top}r\right\rangle \!+\!\log\det\left(2\pi\tilde{C}_{G}\right)\right].
\end{equation}
Write the first term as a trace, cyclically permute the vectors and matrices being multiplied, and use $\left\langle r r^{\top}\right\rangle = C$
to get 
\begin{equation}
\left\langle \log\tilde{\mathcal{L}}\right\rangle =-\frac{1}{2}\left[\Tr\left(\tilde{C}_{G}^{-1}C_{G}\right)\!+\!\log\det\left(2\pi\tilde{C}_{G}\right)\right]. \label{logtildeL}
\end{equation}
The variation of Eq.~\eqref{logtildeL} can be written as
\begin{equation}
\dd\left\langle \log\tilde{\mathcal{L}}\right\rangle
\propto\Tr\left(\tilde{C}_{G} - C_G\right) \dd\tilde{C}_{G}^{-1}. 
\label{dlogtildeL}
\end{equation}
If we had complete freedom to vary $\tilde{C}_G$, we could set
Eq.~\eqref{dlogtildeL} to 0 for all $\dd \tilde{C}_{G}^{-1}$, giving
$\tilde{C}_G=C_G$, which means the optimal approximate model is simply the
one from which the residuals are drawn. However, $\tilde{C}$ is not
allowed to vary freely because we constructed it using a low-rank
approximation. To find an optimal $\Phi$ for $\tilde{C}$, we may
maximize $\left\langle \log\tilde{\mathcal{L}}\right\rangle$ with
respect to $\Phi$.\footnote{An equivalent interpretation of the optimization of $\left\langle \log\tilde{\mathcal{L}}\right\rangle$
in terms of the Kullback-Leibler divergence is discussed in Appendix
\ref{appendA}.}

The details of this maximization are given in Appendices~\ref{append_Phi} and \ref{append_general_Phi}. We show in Appendix~\ref{append_Phi} that $\partial\left\langle \log\tilde{\mathcal{L}}\right\rangle/\partial \Phi=0$ gives a solution
\begin{equation}
    \boxed{\Phi_{\rm MEAL} = F_p^\top D^{-1}C_{\rm gp}D^{-1}F_p, \label{Phi_MEAL}}
\end{equation}
where the subscript ``MEAL'' denotes ``maximizing the ensemble average (log-)likelihood'', and $F_p$ is any basis such that $G^\top F_p$ has the same column span as $G^\top F$, satisfying
\begin{equation}\label{eq:FDF}
    F_p^\top D^{-1}F_p=I.
\end{equation}
Such a basis can be obtained as follows. 

We first define
\begin{align}\label{Pdef}
P\equiv ND^{-1}&=I-M\left(M^\top N^{-1}M\right)^{-1}M^\top N^{-1}\nonumber\\
&= N G (G^\top  N G)^{-1} G^\top ,
\end{align}
which is the
standard generalized least squares projection matrix. Operating with
$P$ annihilates any combination of the columns of $M$, which would be
unobservable.  For a raw basis $F$, such as the Fourier design matrix, we perform a singular value decomposition on the projected basis $PF$,
\begin{equation}
PF=U\Sigma V^{\top}.\label{SVD}
\end{equation}
{Here $U$ consists of an orthonormal projected basis: after the columns in $U$ that correspond to vanishing singular values in $\Sigma$ are removed, we have $PU=U$.}\footnote{{Because floating-point errors make singular values near machine precision unreliable, in practice we discard columns of $U$ whose corresponding singular values are smaller than $10^{-12}$ times the largest singular value in $\Sigma$.}} To further obtain the final basis $F_p$, we apply a linear transformation on $U$. Specifically, we first perform the Cholesky decomposition
\begin{equation}
U^\top D^{-1}U=LL^\top,
\end{equation}
and then define
\begin{equation}
F_p=U(L^\top)^{-1}.
\end{equation}
{It is easy to confirm that $F_p^\top D^{-1}F_p=I$.} Since the timing model is unique for each pulsar, $F_p$ is pulsar-specific. {For a given raw basis $F$, the resulting $F_p$ is not unique; more details are discussed in Appendix~\ref{append_Phi}.}

In summary, starting from a generic basis design matrix $F$ (e.g., Fourier, polynomials, or splines), we construct the pulsar-specific basis $F_p$, and then use Eq.~\eqref{Phi_MEAL} to find the optimal basis-domain covariance $\Phi_{\rm MEAL}$, which enters the PTA likelihood through Eq.~\eqref{eq:Lapprox}. Thus, upgrading from the conventional model to MEAL simply requires the replacements
\begin{equation}
    \boxed{F_{\rm Fourier}\to F_p,\quad
    \Phi_{\rm diag}\to \Phi_{\rm MEAL}.}
\end{equation}
If we fix the white noise parameters, the likelihood Eq.~\eqref{eq:Lapprox} can be further simplified with $F_p$. Note that $r^\top D^{-1} r$ and $\det D$ in the likelihood are irrelevant for Bayesian inference. Defining $w\equiv F_p^\top D^{-1}r$, and using Eqs.~\eqref{eq:L_diagphi} and \eqref{eq:FDF}, we can write
\begin{align}\label{eq:rKr}
    r^T K^{-1} r &= -w^{\top}\left(I+\Phi_{\rm MEAL}\right)^{-1}\Phi_{\rm MEAL} w +\ldots\nonumber \\
    &= w^\top \left(I+\Phi_{\rm MEAL}\right)^{-1}w +\ldots
\end{align}
where we have used the identity $(1+A)^{-1}A = I - (1+A)^{-1}$, and ``$\ldots$'' denotes terms independent of $\Phi_{\rm MEAL}$.  Again using Eq.~\eqref{eq:FDF}, Eq.~\eqref{eq:detK} gives
\begin{equation}\label{eq:ldK}
\log\det K =\log\det\left(I+\Phi_{\rm MEAL}\right) +\ldots.
\end{equation}
Using Eqs.~\eqref{eq:rKr} and \eqref{eq:ldK} in Eq.~\eqref{eq:Lapprox} we find
\begin{empheq}[box=\boxed]{equation}
\begin{aligned}
\label{eq:L_simple}
\log\tilde{\mathcal{L}}=&-\frac{1}{2} w^\top \left(I+\Phi_{\rm MEAL}\right)^{-1}w  \\
&- \frac{1}{2}\log\det\left(I+\Phi_{\rm MEAL}\right) +\ldots.
\end{aligned}
\end{empheq}
We see that the marginalized likelihood $\tilde{\mathcal{L}}$ is now a normal distribution with covariance $I+\Phi_{\rm MEAL}$.

In the next subsection, we describe an efficient numerical procedure for computing $\Phi_{\rm MEAL}$.

\subsection{Numerical computation of $\Phi_{\rm MEAL}$\label{SubIIIC}}

To calculate $\Phi_{\rm MEAL}$, we do not need $C_{\rm gp}$ explicitly. Note that the Wiener-Khinchin theorem can be written as
\begin{equation}
C_{\rm gp}=\int_{0}^{\infty}S(f)\left(r_{\cos} r_{\cos}^{\top}+r_{\sin} r_{\sin}^{\top}\right)\dd f,\label{WK_new}
\end{equation}
where
\begin{equation}
r_{\cos}^{\top}(f)=\left(\cos(2\pi ft_{1}),\cos(2\pi ft_{2}),\ldots\right),
\end{equation}
and similarly for $r_{\sin}^{\top}(f)$. Then, by Eq.~\eqref{Phi_MEAL}, we have
\begin{align}
\Phi_{\rm MEAL}&=\int_{0}^{\infty}S(f)Z(f)\dd f, \label{eq:WKinteg}
\end{align}
where
\begin{align}
Z(f) =& \left(F_{p}^{\top}D^{-1}r_{\cos}\right) \left(F_{p}^{\top}D^{-1}r_{\cos}\right)^\top \nonumber\\
     &+ \left(F_{p}^{\top}D^{-1}r_{\sin}\right) \left(F_{p}^{\top}D^{-1}r_{\sin}\right)^\top,
\end{align}
As $r_{\rm sin}$ and $r_{\rm cos}$ are projected onto the basis space spanned by $F_p$, the dimension of the vectors in the outer products is reduced from $N_{\rm toa}$, typically $\sim 10^5\text{--}10^6$, to the number of basis components $N_F$, typically $\sim 10\text{--}100$.

The next step is to evaluate the integral in Eq.~\eqref{eq:WKinteg} for each element of $\Phi_{\rm MEAL}$. We introduce a uniform frequency grid $\left(f_1,\dots,f_{N_f} \right)$. To capture possible spectral features at frequencies slightly below $1/T$, the lower bound can be set to $f_1\sim 0.01/T\text{--}0.1/T$. In practice, the upper bound is chosen to be a frequency where the white noise is expected to dominate, typically $f_{N_f}\sim 10/T\text{--}100/T$. The grid resolution controls how finely the spectrum can be captured. We define the grid density $g$ such that the total number of grid points is $N_f=\lfloor gT(f_{N_f}-f_1)\rfloor$, so the grid spacing is $\Delta f \approx 1/(gT)$, i.e., $g$ is essentially the number of grid points per interval $1/T$. In practice, we set $g\sim 10$.

To convert the integral \eqref{eq:WKinteg} to a summation, we consider the general case where the integral $\int S(f)\dd f$ does not have a closed form. Standard quadrature methods can be applied. The trapezoidal rule gives
\begin{equation}
\Phi_{\rm MEAL} \approx 
\frac{1}{2}\sum_{i=2}^{{N_f}} 
\big[ S(f_{i-1})Z(f_{i-1})+S(f_{i})Z(f_{i}) \big]\,
\big(f_i-f_{i-1}\big).\label{Phi_MEAL_numerical}
\end{equation}
For a given PSD with fixed $f_1$ and $f_{N_f}$, we can test convergence by increasing the number of grid points $N_f$ (or $g$) and the number of basis functions in $F_p$, i.e., $N_F$, until the inferred parameter posteriors become insensitive to these choices. {When $F_p$ is constructed based on the Fourier basis $F_{\rm Fourier}$, we can set $N_F=2f_{N_f}T$, since further increasing $N_F$ is not expected to capture more features of $\tilde{C}_{\rm gp}$. In this case, we have $N_f\approx gN_F/2$.}

If $\int S(f)\dd f$ has a closed-form expression, more accurate but computationally demanding methods are available. We discuss these alternatives in Appendix~\ref{appendB}.

\subsection{Including pulsar-pair correlations \label{HD}}
We now consider the case that includes inter-pulsar correlations. In addition to a common process shared by all pulsars, a GWB also induces a cross-correlation pattern between any two pulsars. The covariance has the form
\begin{equation}
C_{\rm gw}=\begin{pmatrix}C_{\rm gw}^{(1,1)} & C_{\rm gw}^{(1,2)} & \cdots\\
C_{\rm gw}^{(2,1)} & C_{\rm gw}^{(2,2)} & \cdots\\
\vdots & \vdots & \ddots
\end{pmatrix}.
\end{equation}
The block that describes the correlation of pulsar $a$ and pulsar $b$ can be written as a Wiener-Khinchin integral with a prefactor known as the overlap reduction function (ORF) $\Gamma$,
\begin{equation}
\left(C_{\rm gw}^{(a,b)}\right)_{ij} = \Gamma(\theta_{ab}) \int_{0}^{\infty} \!S_{\rm gw}(f) 
\cos\left[2\pi f\left(t_i^{(a)}-t_j^{(b)}\right)  \right] \, \dd f,
\end{equation}
where $\theta_{ab}$ is the angular separation of the two pulsars, $S_{\rm gw}(f)$ is the PSD of the GWB, and $t_i^{(a)}$ represents the $i$-th TOA of pulsar $a$. Assuming general relativity and an isotropic, unpolarized, and stationary background, the ORF is given by the Hellings-Downs curve \cite{Hellings:1983fr},
\begin{equation}
\Gamma(\theta_{ab})=\frac{3}{2}x\ln(x) - \frac{1}{4}x+\frac{1}{2}+\frac{1}{2}\delta_{ab}, \label{eq:HD}
\end{equation}
where $x=(1-\cos\theta_{ab})/2$, and $\delta_{ab}$ is the Kronecker delta. Note that $\Gamma=1$ for $a=b$.

In the conventional Fourier-diagonal model, the basis design matrix is $F=\left\{F^{(a)}_{\rm Fourier}\right\}$, which has a block-diagonal-like structure. The Fourier-domain covariance for pulsar $a$ and pulsar $b$ is
\begin{equation}
\Phi^{(a,b)}_{\rm diag}=\Gamma(\theta_{\rm ab})\text{diag}\!\begin{pmatrix}\rho_{1} & \rho_{1} & \cdots & \rho_{n_{F}} & \rho_{n_{F}}\end{pmatrix},
\end{equation}
where $\rho_{k}=S_{\rm gw}(k/T)/T$ with $T$ being the whole PTA's observation span.

In the MEAL model, following Eq.~\eqref{Phi_MEAL}, the GWB covariance in the $F_p$-domain is 
\begin{equation}
\boxed{
\Phi^{(a,b)}_{\rm MEAL}
=
F_p^{(a)\top}D_{(a)}^{-1}C^{(a,b)}_{\rm gw}D_{(b)}^{-1}F_p^{(b)}
}
\label{eq:matchL_gw}
\end{equation}
where $F_p^{(a)}$ and $D_{(a)}^{-1}$ are specific to pulsar $a$. The same numerical method described in the previous subsection can be used to compute all correlation blocks. Each diagonal block $\Phi^{(a,a)}_{\rm MEAL}$ includes a PSD for pulsar-specific noise processes (e.g., intrinsic pulsar noise), whereas all blocks, both diagonal and off-diagonal, share a common PSD for the GWB. A single frequency grid shared by all blocks is used for the numerical integration.

In the conventional model, $\Phi^{(a,b)}_{\rm diag}$ can be computed essentially instantaneously, as it is a diagonal matrix scaled by the corresponding ORF value, with each diagonal element given by the PSD evaluated at the corresponding frequency. In contrast, in our model, computing $\Phi^{(a,b)}_{\rm MEAL}$ requires numerical integration for all pulsar-pairs. However, as shown in Appendix~\ref{appendC}, this additional cost does not dominate the overall likelihood evaluation.

\section{Testing with mock data \label{SecIV}}
We test the MEAL model using two mock datasets. The first is a controlled dataset comprising 20 pulsars. A GWB with a power-law PSD is added to white noise. This dataset is designed to test the accuracy of likelihood evaluation in MEAL. The second dataset is a realistic simulation based on the 67 pulsars (and their timing models) in the NANOGrav 15-year dataset. We inject white noise and intrinsic red noise with parameters inferred from the real data, together with a GWB whose PSD is given by a power law plus a Gaussian peak. This dataset serves two purposes: (1) to compare the conventional model and MEAL in terms of parameter inference, and (2) to verify that MEAL correctly recovers the inter-pulsar correlations, specifically the Hellings–Downs curve.

\subsection{Simple dataset: comparing likelihoods \label{simple}}
The first mock dataset is generated from a toy model using \texttt{libstempo}.\footnote{\url{github.com/vallis/libstempo}} It consists of 20 pulsars randomly distributed in the sky, each with an observation span of $T_{\rm obs} = 10~\text{yr}$ and a cadence of 15 days. For simplicity, all TOAs have the same constant white noise level of $1\,\mu\mathrm{s}$, and no intrinsic red noise is included. In addition, a GWB with Hellings-Downs correlations is injected with a power-law PSD
\begin{equation}
S_{\rm crn}(f)=\frac{A_{\rm crn}^{2}}{12\pi^{2}}\left(\frac{f}{1\ \text{yr}^{-1}}\right)^{-\gamma_{\rm crn}}\ \text{yr}^{3},\label{eq:power}
\end{equation}
where $A_{\rm crn}=4\times10^{-15}$ and $\gamma_{\rm crn}=13/3$. The subscript ``crn'' denotes ``common red noise''. 

In this subsection, we calculate the likelihoods using the ``$G$-matrix'' formalism (Eq. \eqref{eq:trueL}). For a power-law GWB, the time-domain covariance $C_{\rm gp}$ can be obtained analytically \cite{vanHaasteren:2008yh}. The small size of this dataset allows us to directly compute the true likelihood, which is to be compared with the MEAL, (oversampled) conventional and FFT-Int approximations. All likelihoods below are evaluated at the injected GWB parameters.

\subsubsection{MEAL vs. (oversampled) conventional model}
We first compare MEAL with the conventional Fourier-diagonal model using $n_F=25$ Fourier modes. For MEAL, the projected basis $F_p$ is constructed from the same $F_{\rm Fourier}$. We set the high-frequency cutoff in Eq.~\eqref{Phi_MEAL_numerical} to $f_{N_f}=25/T_{\rm obs}$, the low-frequency cutoff to $f_1=0.01/T_{\rm obs}$, and the grid density to $g=10$. Thus the two methods have closely matched bases and the same high-frequency cutoff. At the injected GWB parameters, the absolute log-likelihood errors are $0.46$ for the conventional model and $0.0012$ for MEAL. The MEAL covariance thus provides a much more accurate likelihood.

As discussed before, the conventional model approaches the true model as the frequency-grid spacing $\Delta f$ in Eq.~\eqref{eq:DiagPhi} is reduced, because the summation in Eq.~\eqref{eq:discreteWK} then converges to the integral in Eq.~\eqref{eq:WK}. Correspondingly, the basis becomes the oversampled Fourier modes. This provides a natural way to observe how the likelihood converges to the true value. 

\begin{figure}
\centering
\includegraphics[scale=0.5]{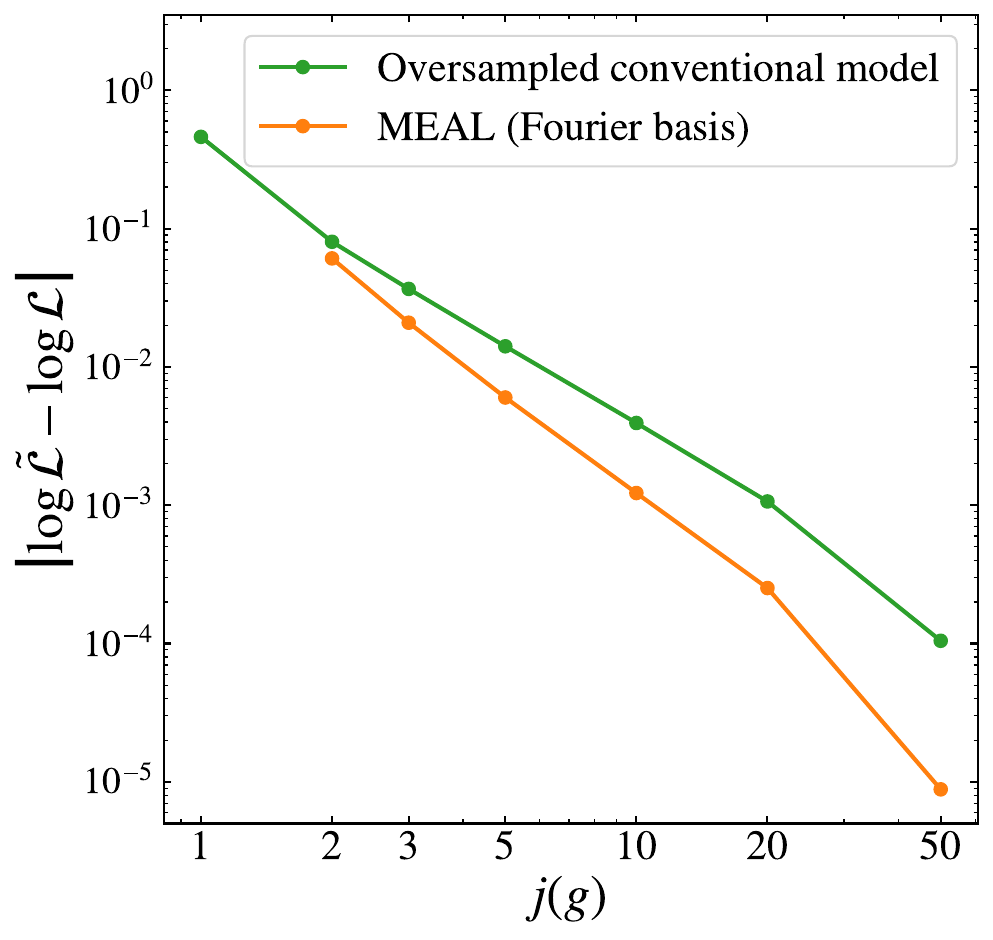}
\caption{\label{fig:oversampledconventional}Absolute log-likelihood error for the oversampled conventional Fourier model and MEAL as a function of their respective frequency grid refinement parameters, $j$ and $g$. Both calculations use the same upper cutoff $f_{\max}=25/T_{\rm obs}$. In the conventional model, $\Delta f=1/(jT_{\rm obs})$ and the number of Fourier modes grows as $n_F=25j$. In MEAL, the projected basis, constructed from $n_F=25$ Fourier modes, is held fixed, while $g$ refines only the numerical Wiener-Khinchin integration used to construct $\Phi_{\rm MEAL}$. Both methods converge toward the true likelihood as the frequency grid is refined, but MEAL achieves smaller errors at comparable grid density without increasing the basis dimension.}
\end{figure}

To compute the likelihood in this oversampled model, we keep an upper bound of $25/T_{\rm obs}$ in Eq.~\eqref{eq:WK}, then in Eq.~(\ref{eq:DiagPhi}), we set $\Delta f = 1/(j T_{\rm obs})$ with $j=1,2,3,5,10,20,$ and $50$. Accordingly, we have $n_F = 25j$ oversampled Fourier modes. Standard PTA analyses correspond to $j=1$. The log-likelihood errors are shown by the green curve in Fig.~\ref{fig:oversampledconventional}. As expected, the approximate likelihood $\mathcal{\tilde L}$ converges to the true one $\mathcal L$ as the frequency grid is refined.

For MEAL, we keep the same $F_p$ and the same high- and low-frequency cutoffs. We then vary only the grid density by setting $g=2,3,5,10,20,$ and $50$. As shown in Fig.~\ref{fig:oversampledconventional}, the MEAL likelihood likewise approaches the true value as the integration grid is refined, and is consistently more accurate than that in the oversampled conventional model.

Although $j$ and $g$ both control the frequency resolution used in the Wiener-Khinchin integral, they play different roles in the two low-rank models. In the conventional model, the basis dimension $2n_F$ grows linearly with $j$, whereas increasing $g$ in MEAL does not change the dimension of $F_p$. Fig. \ref{fig:oversampledconventional} shows that even with as few as $n_F=25$ Fourier modes, MEAL with $g=5$ achieves an accuracy comparable to that of the conventional model that uses $n_F= 250$ ($j= 10$) for an oversampled Fourier basis. Essentially, this is because $\Phi_{\rm MEAL}$ is the projection of the full covariance onto the basis space, whereas the conventional model assigns independent variances to discrete Fourier components. The grid density $g$ controls the numerical accuracy with which the elements of $\Phi_{\rm MEAL}$ are evaluated, while leaving the dimension unchanged. In practice, the choice of $g$ depends on our expectation of the spectral details. 

\subsubsection{MEAL vs. FFT-Int}
As briefly discussed in Subsection \ref{matchingCandC}, an alternative low-rank model, FFT-Int, was recently proposed in Ref. \cite{Crisostomi:2025vue}. In this model, the stochastic process is specified at $N_{\rm node}$ uniformly spaced time-domain nodes and mapped to the TOAs through a linear interpolation basis $F_{\rm int}$. The node spacing, $T_{\rm obs}/(N_{\rm node}-1)$, defines the Nyquist frequency $f_{\rm Nyq}=(N_{\rm node}-1)/(2T_{\rm obs})$. The covariance $\Phi_{\rm FFT}$ between the node amplitudes is obtained directly from the Wiener-Khinchin theorem using an inverse FFT over a frequency grid. The accuracy of the integration is controlled by an integer oversampling parameter $\omega$, which refines the frequency grid to $\Delta f=1/(\omega T_{\rm obs})$ without changing either $F_{\rm int}$ or the dimension of $\Phi_{\rm FFT}$.

\begin{figure}
\centering
\includegraphics[scale=0.5]{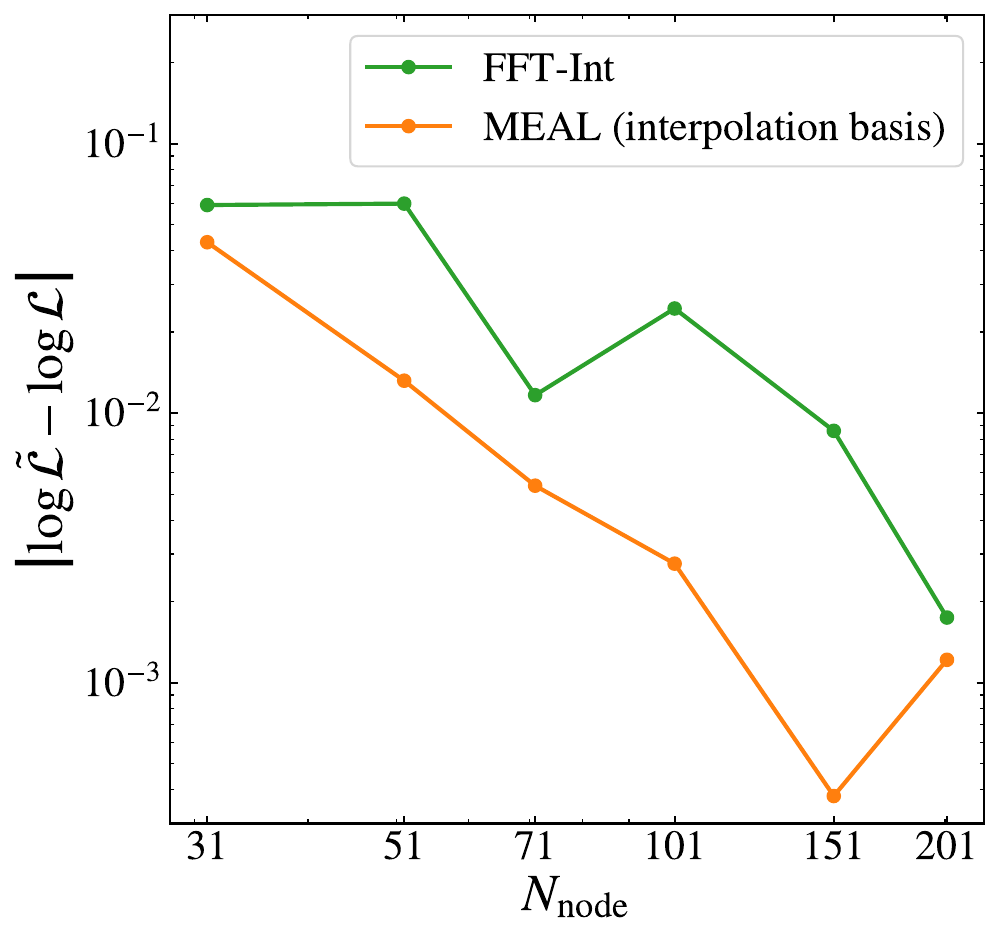}
\caption{\label{fig:FFT-Int}Absolute log-likelihood error for FFT-Int and MEAL as a function of the number of interpolation nodes $N_{\rm node}$. The FFT-Int model uses linear interpolation bases $F_{\rm int}$ of dimension $N_{\rm node}$, while the corresponding MEAL basis $F_p$ is constructed from the same $F_{\rm int}$ at each $N_{\rm node}$. Both methods have the same Wiener-Khinchin frequency range, with $f_{\max}=f_{\rm Nyq}=(N_{\rm node}-1)/(2T_{\rm obs})$. The FFT oversampling parameter $\omega$ and the MEAL grid parameter $g$ are both set to $10$. MEAL gives consistently smaller likelihood errors.}
\end{figure}

We compare MEAL and FFT-Int using interpolation bases $F_{\rm int}$ with dimensions $N_{\rm node}=31, 51, 71, 101,151,$ and $201$. The FFT-Int likelihood errors are shown by the green curve in Fig. \ref{fig:FFT-Int}. The corresponding MEAL likelihood is evaluated using the projected basis $F_p$ constructed from the same $F_{\rm int}$ at each $N_{\rm node}$, so that the two methods use the same interpolation space. We further set $\omega=g=10$ and the same high-frequency cutoff $f_{N_f}=f_{\rm Nyq}$ with $f_{\rm Nyq}T_{\rm obs}=15, 25, 35, 50,75,$ and $100$ in the Wiener-Khinchin calculation. We also impose the same lower cutoff $0.01/T_{\rm obs}$ for both models. As shown in Fig. \ref{fig:FFT-Int}, the MEAL likelihoods are consistently more accurate than the FFT-Int likelihoods.

\subsection{Realistic dataset: comparing parameter recovery \label{realistic}}
The second mock dataset is based on the 67 pulsars in NANOGrav's 15-year data, with an observation span $T_{\rm obs}=16.03~\text{yr}$ \cite{NANOGrav:2023hde}. We take the observed TOAs and the best-fit timing models, and then use \texttt{pta\_replicator}\footnote{\url{https://github.com/bencebecsy/pta_replicator}} to simulate residuals. The injected white noise includes the standard EFAC, EQUAD, and ECORR components for each pulsar \cite{NANOGrav:2023ctt}. The intrinsic pulsar noise for pulsar $a$ is generated by a power-law PSD
\begin{equation}
S_{a}(f)=\frac{A_{a}^{2}}{12\pi^{2}}\left(\frac{f}{1\ \text{yr}^{-1}}\right)^{-\gamma_{a}}\ \text{yr}^{3}.
\end{equation}
The injected white noise and intrinsic red noise parameters come from a representative sample of the fiducial Hellings-Downs analysis of NANOGrav's 15-year data. In addition, we inject a GWB with Hellings-Downs correlations. Its PSD is a power law plus a Gaussian peak (\textsc{PowerLaw+Peak}):
\begin{equation}
S_{\rm gw}(f)=\frac{A_{\rm gw}^{2}}{12\pi^{2}}\!\left(\!\frac{f}{1\ \text{yr}^{-1}}\!\right)^{\!\!\!-\gamma_{\rm gw}}\!\!\!\!\text{yr}^{3}+H_{\rm gw} \exp\!\left[\!-\frac{(f-f_c)^2}{2f_w^2}\!\right],
\label{eq:power2}
\end{equation}
where $A_{\rm gw}=10^{-14.2}$, $\gamma_{\rm gw}=3.2$ (both close to the inferred values in Ref. \cite{NANOGrav:2023gor}), and the Gaussian term models a power excess centered at frequency $f_c$. For widths $f_w$ significantly smaller than $1/T_{\rm obs}$ and center frequencies far from any Fourier frequency $k/T_{\rm obs}$, the conventional Fourier-diagonal model is blind to this feature. In contrast, MEAL includes the effect of the narrow peak on the $k/T_{\rm obs}$ frequencies and so is sensitive to the peak without increasing the dimensionality of $\Phi$.

In the analysis of this dataset, all white noise parameters are fixed to their injected values. The inferred parameters include the amplitudes $A_{a}$ and spectral indices $\gamma_a$ of the intrinsic red noise, as well as the GWB parameters $A_{\rm gw}, \gamma_{\rm gw}, H_{\rm gw}, f_c$ and $f_w$. The priors for these parameters are listed in Table~\ref{tab:injected}.
\begin{table*}[!t]
\begin{center}
\caption{Parameters and priors}
\label{tab:injected}
\begin{tabular}{llll}
\hline\hline
{Parameter}\quad\quad\quad\quad & {Description}\quad\quad\quad\quad   & {Injected value}\quad\quad\quad\quad & {Prior}\\
\hline
\multicolumn{4}{c}{\textbf{\textit{GWB, \textsc{PowerLaw+Peak} Spectrum $(T_{\rm obs}=16.03~\text{yr})$}}} \\[1pt]
$\log_{10}A_{\mathrm{gw}}$& Power-law amplitude & $-14.2$ & Uniform $[-18, -11]$\\
$\gamma_{\rm gw}$& Spectral index & $3.2$ & Uniform $[0, 7]$\\
$\log_{10}H_{\rm gw}~ [\text{s}^3]$& Gaussian amplitude & $-5$ & Uniform $[-9, -4]$\\
$f_c$ [Hz]& Gaussian mean & $10.5/T_{\rm obs}$ & Uniform $[0.1/T_{\rm obs}, 20/T_{\rm obs}]$\\
$\log_{10}f_w$ [Hz]& Gaussian width & $-9.5$& Uniform $[\log_{10}(0.01/T_{\rm obs}), \log_{10}(2/T_{\rm obs})]$\\
\hline
\multicolumn{4}{c}{\textbf{\textit{Intrinsic Red Noise, Power-law Spectrum}}} \\[1pt]
$\log_{10}A_{a}$& Power-law amplitude & ~ & Uniform $[-20, -11]$\\
$\gamma_{a}$& Spectral index& ~ & Uniform $[0, 7]$\\
\hline
\multicolumn{4}{c}{\textbf{\textit{Overlap Reduction Function}}} \\[1pt]
$c_2,c_3,c_4$ & Legendre coefficients &  & Drawn from Dirichlet distribution of order 4\\
\hline
\end{tabular}
\end{center}
\end{table*}
We analyze the dataset using two models:

\begin{enumerate}
\item  The conventional ``$F_{\rm Fourier}~\&~\Phi_{\rm diag}$'' model. The high-frequency cutoff is set to $30/T_{\rm obs}$ for both intrinsic pulsar noise and the GWB. For posterior sampling, we set $\Delta f=1/T_{\rm obs}$. Sampling is performed using \texttt{Discovery}.

\item The MEAL model. The frequency grid for the Wiener-Khinchin integral is set by $f_1=0.01/T_{\rm obs}$, $f_{N_f}=30/T_{\rm obs}$ and $g=10$. The basis $F^{(a)}_p$ for each pulsar is constructed based on the Fourier matrices $F^{(a)}_{\rm Fourier}$ with $n_F=30$ (or $N_F=60$).  In addition, to test the code implementation and its ability to recover inter-pulsar correlations, we do not fix the ORF to the injected Hellings-Downs. Instead, we parameterize it using Legendre polynomials $P_l$ \cite{Gair:2014rwa,NANOGrav:2024tnd}:
\begin{equation}
    ~~~~~~\Gamma(\theta_{ab})=\left(1-\delta_{ab}\right)\sum^{l_{\rm max}}_{l=2}c_l P_l\left(\cos\theta_{ab}\right)+\delta_{ab}.
\end{equation}
The exact Hellings–Downs curve has Legendre coefficients
\begin{equation}
    ~~~~~~c_l=\frac{3}{2}\frac{(l-2)!}{(l+2)!}\left(2l+1\right)
\end{equation}
for $l\ge 2$ and $c_0=c_1=0$. To ensure that the ORF matrix is positive definite, we need 
\begin{equation}
    ~~~~~~c_l\ge0~\text{and}~\sum_l c_l\le1.\label{PD}
\end{equation}
Since the Hellings-Downs curve is well described by the first few terms, in this example we only consider $c_2,c_3$ and $c_4$, i.e., $l_{\rm max}=4$. To satisfy Eq.~\eqref{PD}, we use the Dirichlet distribution to impose priors on the three parameters. Specifically, A four-component vector is drawn from the Dirichlet distribution of order 4; then, three components are assigned to $c_2,c_3$ and $c_4$, respectively. Quantities in the likelihood involving the white noise precision matrix, $D^{-1}$, are precomputed with the help of \texttt{Enterprise}. Likelihood evaluations are implemented in \texttt{JAX} \cite{jax2018github}, and posterior sampling is performed using the Hamiltonian Monte Carlo No-U-Turn Sampler in \texttt{NumPyro} \cite{Phan:2019elc,bingham2019pyro}; both steps are significantly accelerated using modern GPUs.
\end{enumerate}

\begin{figure}[t]
\centering
\includegraphics[scale=0.48]{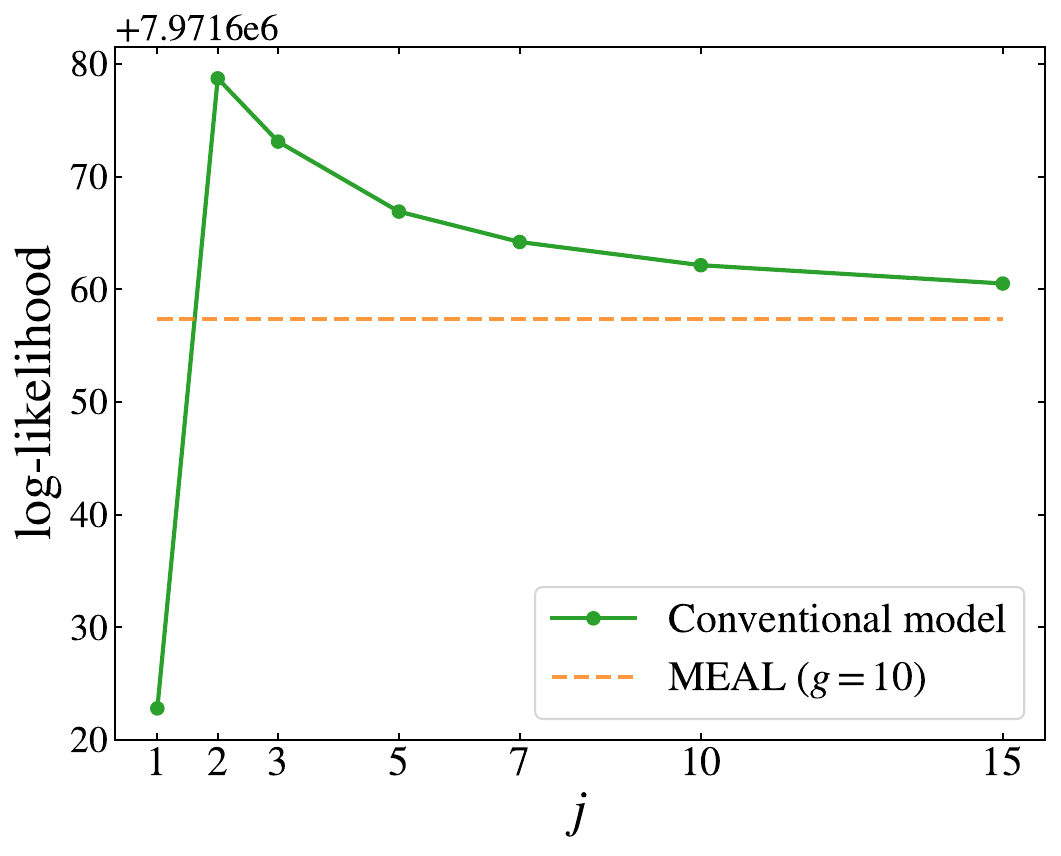}
\caption{\label{fig:likelihood_compare2}Log-likelihoods computed using the oversampled conventional and MEAL models for the realistic dataset. The horizontal axis $j$ is defined through $\Delta f = 1/(jT_{\rm obs})$. The conventional likelihood (solid green) converges to the true one as $j$ increases. The MEAL model (dashed orange) achieves better accuracy with $n_F=30$ (see text).}
\end{figure}

Before presenting the results of Bayesian inference, we first compare the likelihoods in both models using Eq.~\eqref{eq:Lapprox}, evaluated at the injected parameter values. As in the previous subsection, we use $j$, defined through $\Delta f = 1/(jT_{\rm obs})$, to characterize the fineness of the frequency grid used to approximate the Wiener-Khinchin integral in Eq.~\eqref{eq:discreteWK} in the oversampled conventional model. Fig.~\ref{fig:likelihood_compare2}
shows that, as expected, the likelihood from the conventional model (solid green) converges to a constant as $j$ increases. The dashed line represents the MEAL likelihood based on $F_{\rm Fourier}$ with $n_F=30$. This demonstrates once again that MEAL achieves an accuracy comparable to that of the conventional model with a much smaller number of basis functions.

\begin{figure*}[!t]
\centering
\includegraphics[scale=0.44]{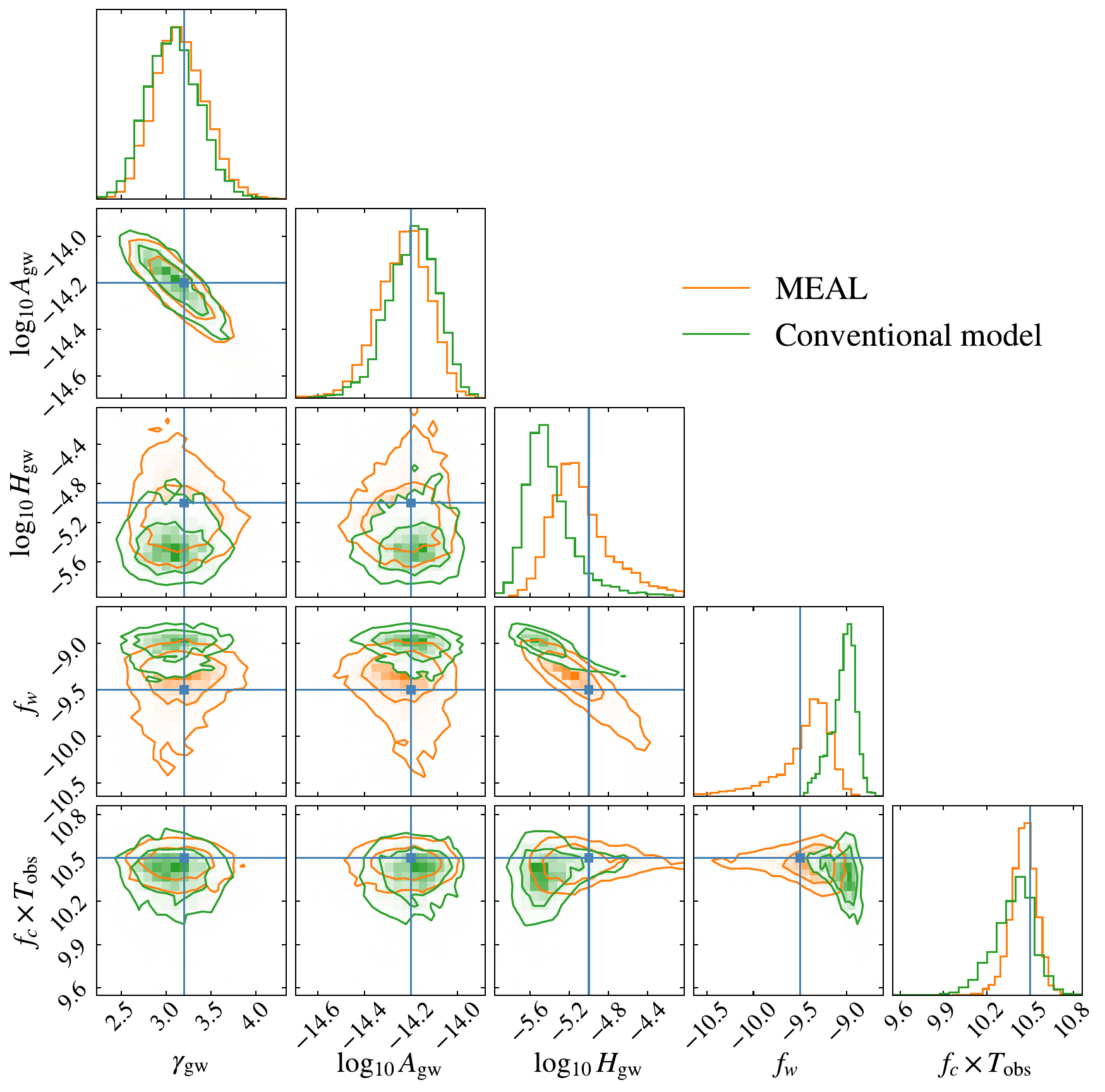}
\caption{\label{fig:MEAL_corner}Corner plot of the GWB parameters in the conventional (green) and MEAL (orange) models. The blue lines represent the injected values.}
\end{figure*}

\begin{figure*}[t]
\centering
\includegraphics[scale=0.52]{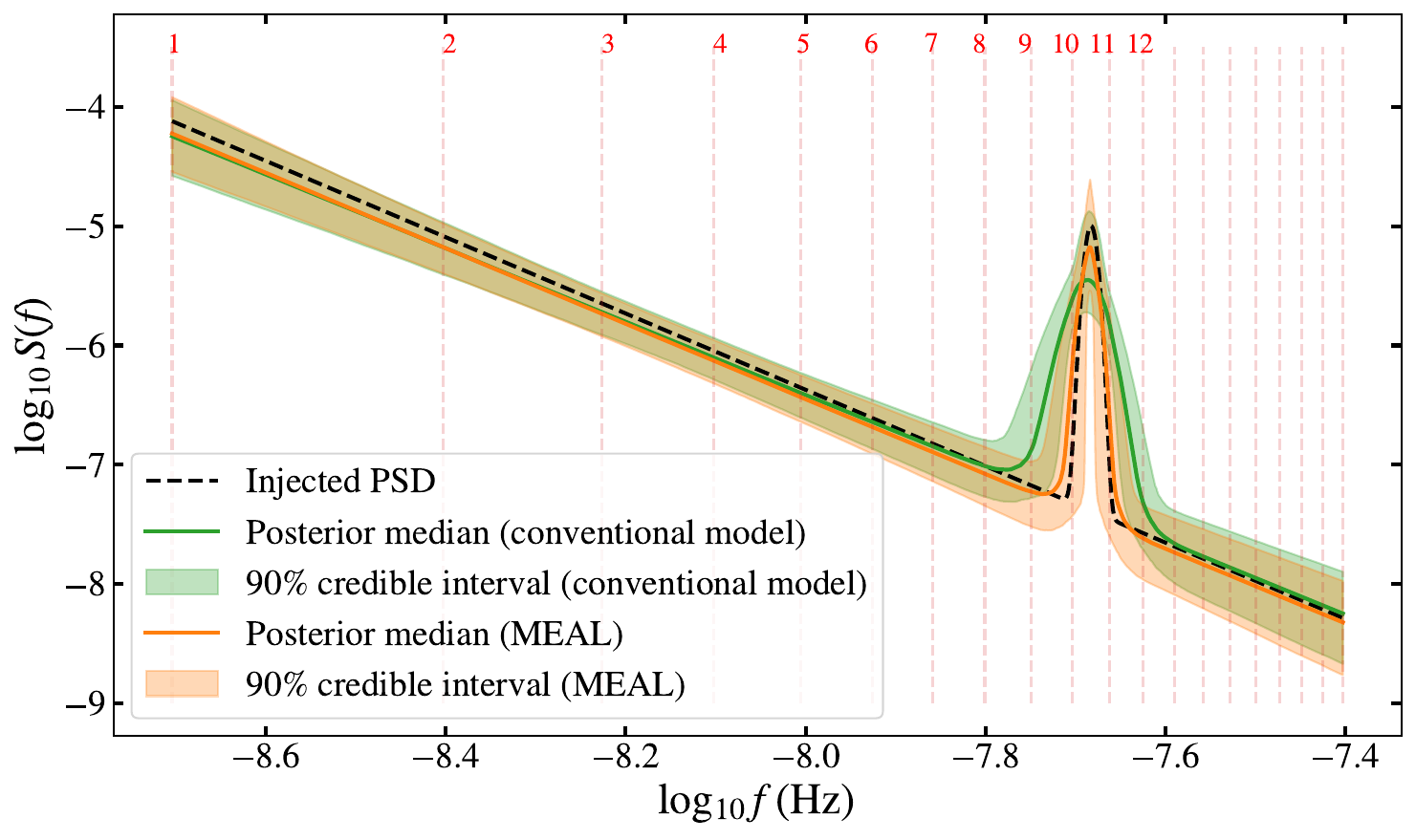}
\caption{\label{fig:MEAL_pSD} Reconstruction of the PSD in the conventional (green) and MEAL (orange) models. The vertical red dashed lines represent the discrete frequencies $k/T_{\rm obs}$ in the conventional model, and the numbers above label $k$. The black dashed curve is the injected PSD, whose Gaussian feature is confined between $10/T_{\rm obs}$ and $11/T_{\rm obs}$. }
\end{figure*}

Fig.~\ref{fig:MEAL_corner}
shows the corner plot of the inferred GWB parameters under the two models. Both models successfully recover the power-law parameters $A_{\rm gw}$ and $\gamma_{\rm gw}$, which is not very surprising because the power law is smooth and well captured by the conventional approach. However, the conventional model is intrinsically limited in its sensitivity to spectral features narrower than a single frequency bin, which in this example is $1/T_{\rm obs}$. At the injected $f_c$ and $f_w$, the peak lies midway between Fourier frequencies and is much narrower than $1/T_{\rm obs}$. If the conventional model is evaluated at the injection parameters, it assigns far too little power to the neighboring frequencies to agree with the injected peak in the data. As a result, although it correctly identifies the central frequency of the Gaussian peak at $f_c \sim 2\times 10^{-8}~\text{Hz}$, the conventional model does not accurately recover either the amplitude or the width. It overestimates the width to make the peak extend across these frequencies so that it is visible, and underestimates the amplitude so the overall power is approximately correct. In contrast, the MEAL model faithfully captures the Gaussian feature. This difference is further illustrated in Fig.~\ref{fig:MEAL_pSD},
which compares the injected PSD with the PSDs reconstructed from posterior samples in the two models.

\begin{figure}[!t]
\centering
\includegraphics[scale=0.34]{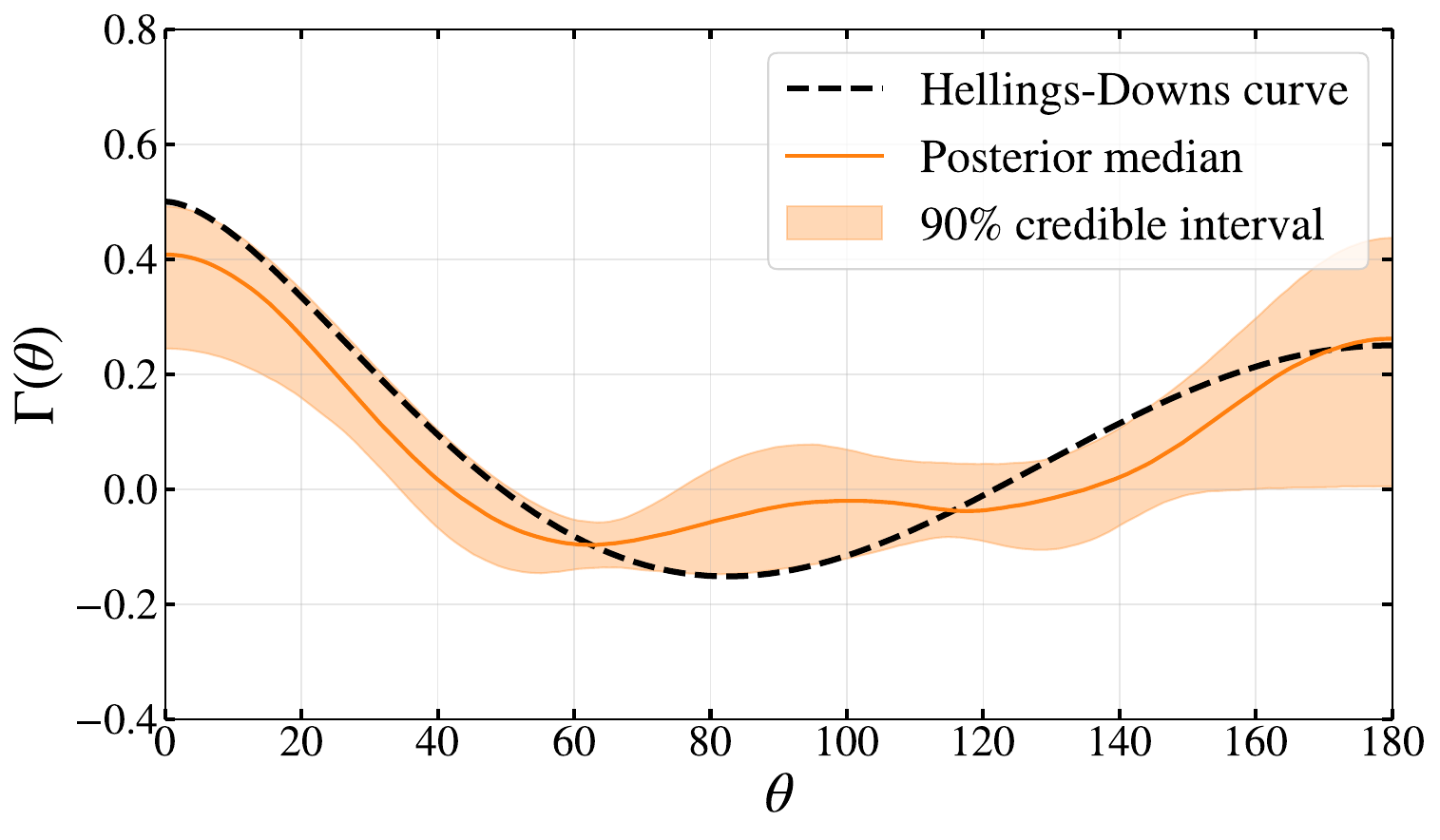}

\caption{\label{fig:MEAL_orf}Reconstruction of the Legendre ORF (up to the fourth order) in the MEAL model. The black dashed curve is the injected Hellings-Downs curve.}
\end{figure}

In addition to the GWB parameters, we simultaneously search for the parameterized ORF in MEAL. In Fig.~\ref{fig:MEAL_orf}, we compare the true ORF, i.e., the Hellings-Downs curve, with the Legendre ORF reconstructed from the posterior samples of the Legendre coefficients $c_2,c_3$ and $c_4$. The $90\%$ credible band is consistent with the injection.



\section{Conclusions and discussion  \label{SecV}}

In this work, we have revisited the modeling of low-frequency processes in PTA, particularly the stochastic GWB, with the goal of improving the accuracy of the PTA likelihood without compromising computational efficiency. The conventional rank-reduced approach, based on a truncated Fourier basis with a diagonal covariance matrix, has played a central role in establishing strong evidence for the nanohertz GWB. However, its discrete formulation has an intrinsic limitation: it is not sensitive to spectral structure on frequency scales smaller than $1/T$, where $T\sim 10~\text{yr}$ is the observation span. As a result, fine-scale spectral features may be incorrectly recovered, potentially leading to biased parameter estimates when the underlying process deviates from a smooth PSD.

We introduced the MEAL (\emph{maximizing ensemble average (log-)likelihood}) model as an alternative. Rather than treating the low-rank approximation as a discretized version of the Wiener-Khinchin integral, we derived the optimal basis-domain covariance $\Phi_{\rm MEAL}$ by maximizing the ensemble average log-likelihood (or, equivalently, minimizing the Kullback–Leibler divergence between the true and approximate likelihoods; see Appendix \ref{appendA}). By construction, the resulting approximate likelihood provides the best statistical description of timing residuals drawn from the ideal model. The MEAL covariance $\Phi_{\rm MEAL}$ incorporates the effects of timing model projection and white noise weighting, and contains a nontrivial off-diagonal structure that encodes information from the continuous PSD without requiring an increased number of basis functions.

Our numerical experiments demonstrate several advantages of the MEAL model over the conventional Fourier-diagonal approximation. In a controlled dataset with a power-law GWB, the MEAL likelihood easily achieves comparable accuracy with far fewer Fourier modes than the conventional model. This improvement comes from the fact that $\Phi_{\rm MEAL}$ captures the full covariance projected onto the basis space, whereas $\Phi_{\rm diag}$ contains only the PSD values assigned independently to the discrete Fourier modes. Using a realistic simulated dataset, we showed that MEAL accurately recovers spectral features that the conventional approach cannot. In particular, narrow-band structures in the PSD, such as a Gaussian peak, can be faithfully reconstructed. We further verified that MEAL correctly recovers inter-pulsar correlations (the Hellings–Downs curve in our test), demonstrating its robustness for full PTA analyses.

From a computational perspective, the MEAL model preserves the advantages of a low-rank formulation. Although the construction of $\Phi_{\rm MEAL}$ introduces an additional numerical integration step, the cost does not dominate the likelihood evaluation (see Appendix~\ref{appendC}) and can be efficiently implemented, especially with modern computational tools such as \texttt{JAX} and GPU acceleration. As a result, MEAL offers a practical upgrade to existing analysis pipelines.

MEAL's sensitivity to spectral features makes it particularly well suited for searches for non-power-law signals or scenarios involving a small number of loud sources. More broadly, the MEAL construction can be viewed as a variational approximation to the true likelihood, suggesting extensions to non-Gaussian processes and hierarchical population models, where one seeks an optimal low-dimensional representation of more complex signals.

In summary, the MEAL model provides a statistically well-motivated and computationally efficient method for improving likelihood accuracy in PTA analyses. By moving beyond diagonal approximations and directly optimizing the likelihood, it leads to faithful inference of both spectral properties and spatial correlations of the GWB, and thus has the potential to play an important role in the next generation of PTA discoveries.

\section*{Acknowledgments}
We thank Yuri Levin for helpful discussions. We also acknowledge the use of Google's Gemini and OpenAI's ChatGPT for assistance with code development and debugging. HD was supported by Yuri Levin’s Simons Investigator Grant PG012519. KDO\ was supported in part by NSF Grant Nos.~2020265 and 2207267.

\appendix

\section{Equivalent interpretation of MEAL: matching the likelihoods using KL divergence \label{appendA}}

There is an equivalent perspective on the maximization of the ensemble average log-likelihood. Since the likelihood can be interpreted as a probability distribution, we may seek to minimize the discrepancy between the true and approximate likelihoods. In other words, we want to know which $\Phi$ gives a distribution as close as possible to the one that generates the residuals. A standard measure for the ``distance'' between two distributions is the Kullback–Leibler (KL) divergence. For distributions $P$ and $Q$ with probability densities $p$ and $q$ respectively, the KL divergence is
\begin{equation}
D_{\text{KL}}(P\|Q)\equiv\int p(x)\log\frac{p(x)}{q(x)}\dd x,
\end{equation}
which quantifies the information lost when we use $Q$ to approximate $P$. 

In our case, both the true and approximate likelihoods are multivariate normal distributions. Let $n_{C_G}$ be the dimension of $C_G$. The KL divergence is\footnote{If $\tilde{C}_{G}=C_{G}$, then $D_{\text{KL}}=0$
and the two distributions coincide.}
\begin{equation}
D_{\text{KL}}=\frac{1}{2}\left[\Tr\left(\tilde{C}_{G}^{-1}C_{G}\right)-n_{C_G}+\log\!\left(\frac{\det\tilde{C}_{G}}{\det C_{G}}\right)\right],
\end{equation}
which, up to terms independent of $\tilde{C}$ (and thus $\Phi$), is simply $-\left\langle \log\tilde{\mathcal{L}}\right\rangle$ (Eq.~\eqref{logtildeL}). Therefore, minimizing the KL divergence is equivalent to maximizing the ensemble average log-likelihood of the approximate model.

\section{MEAL solution through whitening and timing model projection \label{append_Phi}}
To maximize $\left\langle \log\tilde{\mathcal{L}}\right\rangle$ with respect to $\Phi$, it is instructive to rewrite the likelihood by whitening the variables and isolating the subspace orthogonal to the timing model. To this end, we first define the variables in the $G$-projected space:
\begin{align}
    r_G &= G^\top r,\\
    F_G &=G^\top F,\\
    N_G &=G^\top N G,\\
    C_{\text{gp},G} &=G^\top C_{\rm gp} G.\label{eq:CgpG}
\end{align}
The full covariance in this space is $C_G=N_G+C_{\text{gp},G}$. We then define the whitening matrix as $N_G^{-1/2}$, which is used to whiten the projected variables:
\begin{align}
    r_W&=N_G^{-1/2} r_G,\\
        F_W&=N_G^{-1/2} F_G,\\
C_{\text{gp},W}&=N_G^{-1/2} C_{\text{gp},G} N_G^{-1/2}.\label{eq:CgpW}
\end{align}
Then the full covariance and the approximate covariance in this whitened space simplify to 
\begin{align}
    C_W&=I+C_{\text{gp},W},\\
    \tilde{C}_W&=I+F_W\Phi F_W^\top,\label{btC}
\end{align}
respectively, where the white noise matrix has been reduced to an identity matrix.

The effect of the basis depends only on the space spanned by the basis vectors.  Thus, to simplify later calculations, we can construct an orthonormal basis $F_W^\prime$ for the column span of $F_W$, for example using singular value decomposition,
\begin{equation}
    F_W=F_W^\prime S W^\top . \label{F_W}
\end{equation}
Replacing $F_W$ with $F_W^\prime$ in the likelihood gives
\begin{equation}
\left\langle \log\tilde{\mathcal{L}}\right\rangle  
=-\frac{1}{2}\left[\Tr\left(\tilde{C}_W^{-1}C_W\right)\!
+\!\log\det\left(2\pi N_G \tilde{C}_W\right)\right], \label{logtildeL2}
\end{equation}
with variation
\begin{equation}
\dd\left\langle \log\tilde{\mathcal{L}}\right\rangle
\propto \Tr\left(\tilde{C}_W - C_W\right) \dd\tilde{C}_W^{-1}. 
\label{dlogtildeL2}
\end{equation}
From Eq.~(\ref{btC}) we observe that $\tilde{C}_W F_W' =
F'_W (I+\Phi)$
and thus $\tilde{C}_W^{-1} F_W' = F_W' (I+\Phi)^{-1}$.  Then
\begin{align}
\dd\tilde{C}_W^{-1} &= - \tilde{C}_W^{-1} \dd\tilde{C}_W
\tilde{C}_W^{-1}\nonumber\\
&= - \tilde{C}_W^{-1} F'_W \dd\Phi F_W'^{T}\tilde{C}_W^{-1}\nonumber\\
&= -F'_W (I+\Phi)^{-1} \dd\Phi (I+\Phi)^{-1} F'^\top _W.
\end{align}
Putting this in Eq.~(\ref{dlogtildeL2}), permuting the trace, and observing that
$(I+\Phi)^{-1} \dd\Phi (I+\Phi)^{-1}$ can take any value, we find
\begin{equation}
0 = F_W'^\top  \left(\tilde{C}_W - C_W\right) F_W'
= \Phi - F_W'^\top C_{\text{gp},W}F_W'
\end{equation}
and thus
\begin{equation}
\Phi=F_W'^\top C_{\text{gp},W}F_W'.
\label{eq:matchL_}
\end{equation}

Directly using Eq.~(\ref{eq:matchL_}) to evaluate $\Phi$ is
computationally prohibitive due to the need to explicitly form $G$ and
$N_G$. Instead, let us find some basis $F_p$ in the original coordinate system such that $G^\top  F_p$ has the same column span as $G^\top  F$ and Eq.~\eqref{eq:FDF} holds.  Then we can set
\begin{equation}\label{eq:newFW}
    F_W' = N_G^{-1/2} G^\top  F_p.
\end{equation}
This has the same span as $F_W$ above, and gives
$F_W'^\top  F_W' = F_p^\top  G N_G^{-1 }G^\top  F_p = F_p^\top  D^{-1} F_p = I$.  These are the properties we used above, so $\Phi$ is given by Eq.~(\ref{eq:matchL_}).  Using Eqs.~\eqref{eq:newFW}, \eqref{eq:CgpG}, and \eqref{eq:CgpW}, this becomes
\begin{equation}
    \Phi = F_p^\top D^{-1}C_{\rm gp}D^{-1}F_p,
\end{equation}
which is our Eq.~\eqref{Phi_MEAL}.

\section{Alternative MEAL derivations\label{append_general_Phi}}

\subsection{Using generalized singular value decomposition}
As an alternative to the procedure for finding $F_p$ in the main text, we can use generalized singular value decomposition (GSVD)\footnote{There are two unrelated ideas called GSVD.  We mean here the procedure in which weighting matrices appear in the equations to be solved \cite{toussile2024singular}.}.  Using $P$ defined in Eq.~(\ref{Pdef}), we look for a decomposition
\begin{equation}
P F = F_p S W^\top \label{PFGSVD}
\end{equation}
where
\begin{equation}\label{PFGVSD2}
F_p^\top  N^{-1} F_p = I.
\end{equation}
From Eq.~(\ref{Pdef}) we see that $G^\top  P = G^\top $. Applying $G^\top $ on the left to Eq.~(\ref{PFGSVD}) shows that $G^\top  F_p$ has the same column span as $G^\top  F$.  Furthermore since $D^{-1}P = N^{-1}P$, $F_p$ satisfies Eq.~\eqref{eq:FDF}.

To solve Eqs.~(\ref{PFGSVD},\ref{PFGVSD2}) without creating the matrix $N^{-1/2}$, we can diagonalize $F^\top  P^\top N^{-1} P F = W S^2 W^\top $, with $S$ diagonal and $W$ orthogonal.  Then we write $F_p = P F W S^{-1}$, and Eqs.~(\ref{PFGSVD},\ref{PFGVSD2}) are satisfied.  As in the main text, to improve numerical stability we may wish to remove small eigenvalues from $S^2$ and the corresponding columns from $W$.

\subsection{Using Moore-Penrose inverse}
The maximization of the ensemble average log-likelihood can result in a solution similar in form to Eq. \eqref{eq:matchC}. To begin with, the variation of the ensemble average log-likelihood is given by Eq.~\eqref{dlogtildeL}, which can be explicitly written as
\begin{align}
&\dd\left\langle \log\tilde{\mathcal{L}}\right\rangle \\\propto& \Tr\left(C_{G}-\tilde{C}_{G}\right)\tilde{C}^{-1}_{G}\dd\tilde{C}_{G}\tilde{C}^{-1}_{G}\nonumber\\
=&\Tr\tilde{C}^{-1}_{G}\left(C_{G}-\tilde{C}_{G}\right)\tilde{C}^{-1}_{G}\dd\tilde{C}_{G}\\
=&\Tr F^\top G\tilde{C}^{-1}_{G} G^\top\left(C_{\rm gp}-F \Phi F^\top\right)G\tilde{C}^{-1}_{G}G^\top F \dd\Phi. \nonumber\label{dlogtildeL_2}
\end{align}
Here, $F^\top G\tilde{C}_G^{-1}G^\top$ can be further written as
\begin{align}
& ~F^\top G\tilde{C}_G^{-1}G^\top \nonumber\\ 
=& ~F^\top G\left[G^{\top}\left(N+F\Phi F^{\top}\right)G\right]^{-1}G^\top\nonumber\\
=& ~F^\top D^{-1}-F^\top D^{-1}F\left(I+\Phi F^{\top}D^{-1}F\right)^{-1}\Phi F^{\top}D^{-1}\nonumber\\
=&\left[I-F^\top D^{-1}F\left(I+\Phi F^{\top}D^{-1}F\right)^{-1}\Phi \right]F^{\top}D^{-1}\nonumber\\
=& \left(I+F^\top D^{-1}F \Phi\right)^{-1}F^{\top}D^{-1},
\end{align}
and, similarly,
\begin{equation}
G\tilde{C}_G^{-1}G^\top F   = D^{-1}F\left(I+\Phi F^{\top}D^{-1}F\right)^{-1},
\end{equation}
where we have used $D^{-1}=G(G^\top NG)^{-1}G^\top$ and a variation of the Woodbury matrix identity: $(A+UBV)^{-1}=A^{-1}-A^{-1}U(I+BVA^{-1}U)^{-1}BVA^{-1}$.
Inserting these two expressions into Eq. (\ref{dlogtildeL_2}), the stationary condition $\partial\left\langle \log\tilde{\mathcal{L}}\right\rangle/\partial \Phi=0$ gives the matrix equation
\begin{equation}
\left(F^{\top}D^{-1}F\right)\Phi\left(F^{\top}D^{-1}F\right)=F^{\top}D^{-1}C_{\text{gp}}D^{-1}F. \label{MEAL_general_equation}
\end{equation}
A solution to this system is
\begin{equation}
\Phi=\left(F^{\top}D^{-1}F\right)^+ \left(F^{\top}D^{-1}C_{\text{gp}}D^{-1}F\right)\left(F^{\top}D^{-1}F\right)^+, \label{MEAL_generalized}
\end{equation}
where the plus sign denotes the Moore-Penrose inverse (or pseudoinverse).  To show that this is indeed an exact solution, we first note that $F^\top D^{-1} F=F^\top D^{-1} N D^{-1} F$. By Observation 7.1.8 in Ref. \cite{horn2012matrix}, because the white noise covariance $N$ is positive definite, we have
\begin{equation}
    \text{rank}\left(F^\top D^{-1} N D^{-1} F\right)=\text{rank}\left(F^\top D^{-1}\right).
\end{equation}
By Corollary 4.2.3 in Ref. \cite{harville1998matrix}, the column space of $F^\top D^{-1} F$ is a subspace of the column space of $F^\top D^{-1}$:
\begin{equation}
    \text{col}\left(F^\top D^{-1} F\right)\subseteq \text{col}\left(F^\top D^{-1}\right).    
\end{equation}
Then it follows that\begin{equation}
    \text{col}\left(F^\top D^{-1} F\right)= \text{col}\left(F^\top D^{-1}\right).    
\end{equation} 
Likewise, this column space is also shared by $F^\top D^{-1} C_{\rm gp} D^{-1} F$. By the definition of Moore-Penrose inverse, $\left(F^\top D^{-1} F\right)\left(F^\top D^{-1} F\right)^{+}$ is a projection onto the column space of $F^\top D^{-1} F$. We thus have
\begin{align}
    &\left(F^\top D^{-1} F\right)\left(F^\top D^{-1} F\right)^{+}\left(F^\top D^{-1} C_{\rm gp} D^{-1} F\right)\nonumber\\
    =&F^\top D^{-1} C_{\rm gp} D^{-1} F.
\end{align} 
Then it immediately follows that Eq. (\ref{MEAL_generalized}) is an exact solution of Eq. (\ref{MEAL_general_equation}). 

In practice, however, we do not use this solution for implementation because the Moore-Penrose inverse is possibly ill-conditioned and is thus relatively unstable. This is why, before solving for $\Phi$, we constructed a better-behaved projected basis $F_p$ from the raw basis $F$.

\section{More accurate computation of Wiener-Khinchin integral \label{appendB}}
 Rather than using numerical integration methods such as the trapezoidal rule and Simpson's rule, the Wiener-Khinchin integral (\ref{eq:WKinteg}) can be computed more accurately if $\int S(f)\dd f$, $\int fS(f)\dd f$, etc., have closed-form expressions. We want to compute $\int S(f)Z(f)\dd f$ using a frequency grid $\{f_i\}$. After obtaining $Z(f_i)$ at the grid points, we can use a spline to approximate $Z(f)$ at $f_i < f < f_{i+1}$. For instance, using linear interpolation, we have
\begin{align}
Z(f) &\approx Z(f_i) 
      + \frac{Z(f_{i+1}) - Z(f_i)}{f_{i+1} - f_i}(f - f_i).
\label{eq:Zinterp}
\end{align}
Then for the $i$-th grid segment, we have
\begin{align}
\int_{f_i}^{f_{i+1}}\! S(f)Z(f)\dd f 
&\approx 
\frac{f_{i+1}Z(f_{i})\! -\! f_iZ(f_{i+1})}{f_{i+1}\! -\! f_i} 
   \int_{f_i}^{f_{i+1}} \!S(f)\dd f  \nonumber \\
&\quad + \frac{Z(f_{i+1})\! -\! Z(f_i)}{f_{i+1}\! -\! f_i} 
   \int_{f_i}^{f_{i+1}}\! fS(f)\dd f.
\label{eq:bin_integral}
\end{align}
Thus, we need to compute two integrals for each segment. This is particularly useful when they can be evaluated analytically (such as for a power law). We then sum the results for all segments to get $\int S(f)Z(f)\dd f$. Similarly, if we want to use a higher-order spline, we need to compute $\int S(f)\dd f$, $\int fS(f)\dd f$, $\int f^2 S(f)\dd f$, etc., for each grid segment.

\section{Analysis of computational cost \label{appendC}}
In this appendix, we estimate the computational cost of likelihood evaluation in our model and compare it with that of the conventional model. We focus on the case that includes pulsar-pair correlations. Following common practice, we assume that all white noise parameters are fixed from individual pulsar analyses. Consequently, quantities involving the white noise covariance, such as $F^\top D^{-1} r$ and $Z(f_i)$, can be precomputed.

In the ``$F_{\rm Fourier}~\&~\Phi_{\rm diag}$'' model, the computational bottleneck in likelihood evaluation is the Cholesky decomposition in $\Theta = \left(\Phi^{-1}+F^\top D^{-1} F\right)^{-1}$, which requires (in the limit of a large matrix) $N_F^3 N_{\rm psr}^3/3$ floating point operations. Here $N_F$ is the number of columns of $F$ and $N_{\rm psr}$ is the number of pulsars in the array. In the MEAL model, one needs to decompose $I+\Phi$ (see Eq.~\eqref{eq:L_simple}), with the same computational requirement.

The MEAL model has an additional cost from constructing $\Phi_{\rm MEAL}$ using numerical integration. Let $N_f$ be the number of frequency grid points used in the integration. The cost of computing $\Phi_{\rm MEAL}$ following Eq.~(\ref{Phi_MEAL_numerical}) is then $N_f N_F^2 N_{\rm psr}^2$ operations. Thus the decomposition in $\Theta$ is dominant for sufficiently large $N_{\rm psr}$. However, we are not always in that limit.

The value of $N_f$ depends on the upper bound of the frequency grid and the frequency-grid density needed to capture the structure of the PSD. For a simple power-law PSD, $N_f$ can be as small as 100 or even 10, comparable to $N_F$.  In such cases, the computation of $\Phi_{\rm MEAL}$ adds little to the total time. For more complex spectra, such as the \textsc{PowerLaw+Peak} PSD considered in our experiments, one may require $N_f$ to be as large as 1000,\footnote{For example, as discussed in Subsection~\ref{SubIIIC}, when $F_p$ is constructed based on the Fourier basis, fully resolving the structure of $\tilde{C}_{\rm gp}$ requires $N_f\approx gN_F/2$. For $N_F=100$ and $g=20$, this gives $N_f\approx 1000$.} which is only somewhat less than a typical $N_F N_{\rm psr}/3$.  While highly optimized routines exist for Cholesky decomposition, the addition of matrices and multiplication by a scalar in Eq.~(\ref{Phi_MEAL_numerical}) is also very rapid.  In general, we do not expect the additional step of numerical integration in our procedure to significantly increase the likelihood computation time.

\begin{figure}
\centering
\includegraphics[scale=0.41]{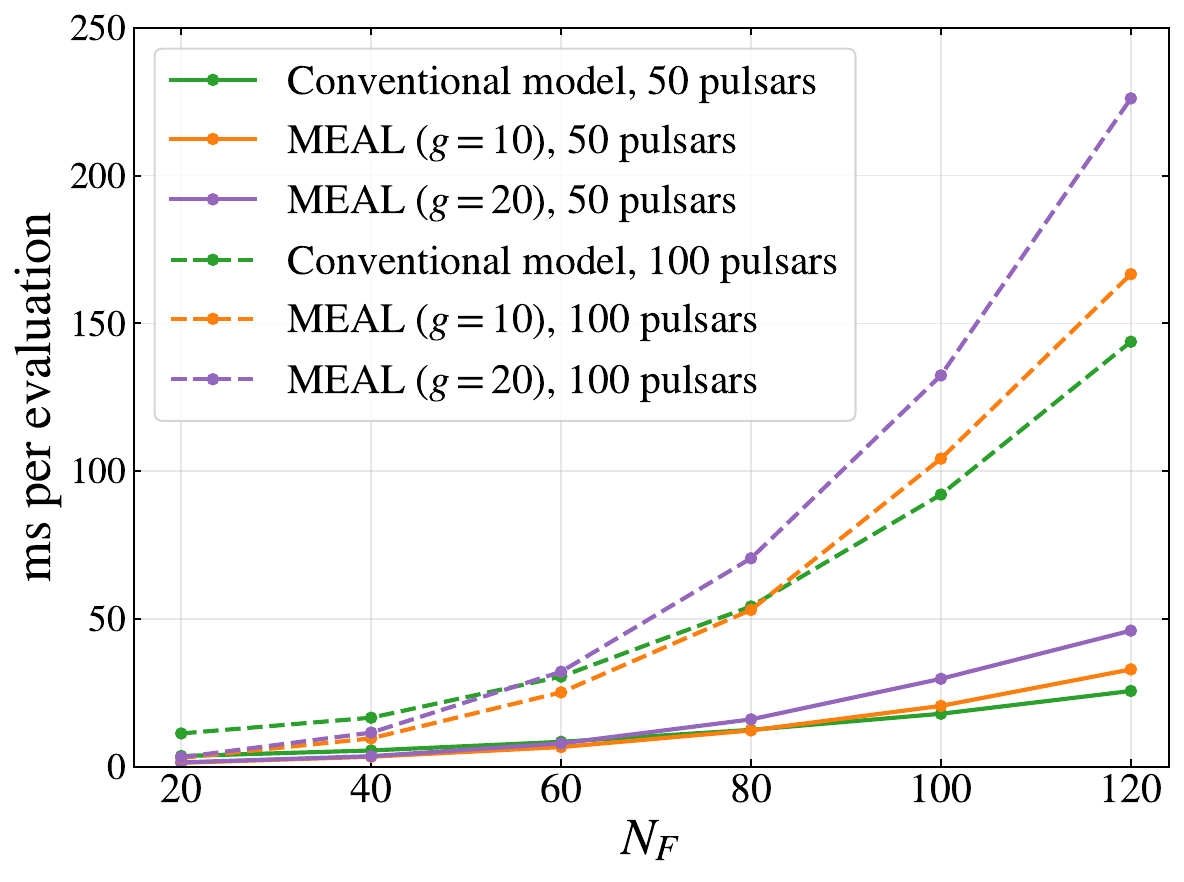}

\caption{\label{fig:likelihood_speed} Average likelihood evaluation times for the conventional and MEAL models, using two simple datasets similar to the dataset in Subsection~\ref{simple}. The horizontal axis represents the number of basis functions, and $g$ denotes the density of the frequency grid for numerical integration in MEAL.}
\end{figure}

To verify this, we benchmark the likelihood evaluation using a simple dataset similar to the one in Subsection~\ref{simple}. We simulate 100 pulsars with an 
 observation span of 20 years and a cadence of 15 days. We evaluate Eq.~(\ref{eq:Lapprox}) for the conventional model (using \texttt{Discovery}) and Eq.~(\ref{eq:L_simple}) for MEAL (using our own code), varying the number of basis components $N_F$.\footnote{Note that we are comparing two likelihoods from their respective implementations. We believe this comparison is reasonable because the major computational step in both cases is the Cholesky decomposition of an $N_F\times N_F$ matrix, which is carried out using the \texttt{jax.scipy} module in both implementations.} In both models, the likelihood is computed for power-law intrinsic pulsar noise and a power-law GWB, with the ORF fixed to the injected Hellings-Downs. To examine the impact of $N_{\rm psr}$, we also analyze a sub-dataset containing 50 of the 100 pulsars. The runtime per likelihood evaluation is averaged over 1000 random samples.\footnote{The GPU used for this benchmark is NVIDIA Tesla V100S (32 GB HBM2).} The results are shown in Fig.~\ref{fig:likelihood_speed}. As expected, the computational cost of the MEAL likelihood is comparable to that of the conventional approach. For the 50-pulsar sub-dataset, even in the regime with a large number of basis components (e.g., $N_F=120$) and a dense frequency grid for numerical integration (e.g., $g=20$), the evaluation time of the MEAL likelihood increases by only a factor of about 2. For the full 100-pulsar dataset, in the large-$N_F$ regime, the relative difference between the two models becomes even smaller, since $N_F N_{\rm psr}/3$ becomes even more dominant over $N_f$, making the additional cost in MEAL less significant.

Lastly, note that, as in the conventional model, the basis coefficients are analytically marginalized in MEAL. Recently developed coefficient-sampling methods, such as \texttt{Prometheus} \cite{Gundersen:2026ows}, can have substantially different computational characteristics, and the additional operations required by MEAL may become more significant in this setting. Optimizing MEAL for such methods is left for future work.

\section{Including chromatic noise \label{appendD}}
Besides the GWB and intrinsic pulsar noise, there are other noise components in PTA data that can be modeled by Gaussian processes. A prominent example is chromatic noise caused by, e.g., dispersion measure (DM), scattering, and solar wind, whose induced timing delay is radio-frequency dependent. Accurately modeling chromatic noise is essential for PTA analyses, as they can otherwise mimic signals such as the GWB.

For example, the time delay of a given radio frequency $\nu$ induced by DM variations scales as $\delta t \propto \nu^{-2}$. When modeling DM as a stationary Gaussian process, the covariance of the timing residuals is given by
\begin{equation}
    \left(C_{\text{DM}}\right)_{ij}\propto (\nu_i \nu_j)^{-2}\int_{0}^{\infty}\!S_{\rm DM}(f)\cos\!\left[2\pi f(t_{i}-t_{j})\right]\dd f,\label{eq:WK_DM}
\end{equation}
where $S_{\rm DM}(f)$ is the PSD that characterizes the stochastic process, typically a power law. Similar to Eq.~(\ref{WK_new}), this can be rewritten as
\begin{equation}
C_{\text{DM}}\propto \int_{0}^{\infty}S_{\rm DM}(f)\left(r^{\rm DM}_{\cos} r_{\cos}^{\rm DM\top}+r^{\rm DM}_{\sin} r_{\sin}^{\rm DM\top}\right)\dd f,
\end{equation}
where
\begin{equation}
r_{\cos}^{\rm DM}(f)=\left(\frac{\cos(2\pi ft_{1})}{\nu_1^2},\frac{\cos(2\pi ft_{2})}{\nu^2_2},\ldots\right)^\top,
\end{equation}
and similarly for $r_{\sin}^{\rm DM}(f)$. Then we can follow Eqs. (\ref{eq:WKinteg}-\ref{Phi_MEAL_numerical}) and obtain the DM's covariance in the basis space, $\Phi^{\rm DM}_{\rm MEAL}$, for each pulsar.

Higher-order chromatic processes, such as interstellar scattering, are sometimes modeled with a varying radio-frequency index \cite{Larsen:2026pru}. In this case, quantities such as $F_p^\top D^{-1} r_{\cos}$ in $\Phi_{\rm MEAL}$ must be recomputed as the index varies during sampling, which increases the computational cost. In conventional low-rank methods, the radio frequency $\nu$ enters through the basis, whereas in MEAL it enters through the construction of $\Phi_{\rm MEAL}$.

It should be noted that MEAL applies to chromatic noise only when the process is modeled using a specific PSD. Unlike the GWB, whose PSD is often motivated by astrophysical or cosmological theories, there is no well-established, physically motivated parameterization for chromatic processes. Therefore, one may choose to model the covariance not through the Wiener-Khinchin theorem, but through Gaussian process kernels. These flexible statistical models serve as phenomenological descriptions of chromatic noise and can be favored by the data for some pulsars \cite{Hazboun:2025syp,Larsen:2026pru,Agarwal:2026dlv}. 

\bibliographystyle{apsrev4-1}   
\bibliography{MEAL_references}

\end{document}